\documentclass[journal,comsoc]{IEEEtran}
\usepackage{subcaption}
\usepackage{graphicx}
\usepackage{amsfonts}
\usepackage{amssymb}
\usepackage{amsmath}
\usepackage{amsthm}
\usepackage{cite}
\usepackage{mathrsfs}
\usepackage[displaymath,mathlines]{lineno}
\usepackage{color}
\usepackage{tabulary}
\usepackage{multirow}
\usepackage{algpseudocode}
\usepackage{algorithm,algpseudocode}
\usepackage{algorithmicx}
\usepackage{pbox}
\usepackage{multicol}
\usepackage{lipsum}
\usepackage{amsthm}
\usepackage{relsize}
\usepackage{lipsum}
\usepackage{epstopdf}
\usepackage{mathtools}
\usepackage{booktabs}
\usepackage{comment}
\usepackage{amssymb}
\usepackage{stackrel}
\usepackage[dvipsnames]{xcolor}
\newcolumntype{P}[1]{>{\centering\arraybackslash}p{#1}}
\usepackage{calc}% http://ctan.org/pkg/calc
\usepackage{makecell}
\usepackage{cases}

\allowdisplaybreaks

\makeatletter
\newcommand{\doublewidetilde}[1]{{%
		\mathpalette\double@widetilde{#1}}}
\newcommand{\double@widetilde}[2]{%
	\sbox\z@{$\m@th#1\widetilde{#2}$}%
	\ht\z@=.5\ht\z@
	\widetilde{\box\z@}}
\makeatother

\usepackage{accents}

\newtheorem{lemma}{Lemma}

\newtheorem{remark}{Remark}

\begin{document}

\title{\huge Double RIS-Assisted MIMO Systems Over Spatially Correlated Rician Fading Channels and Finite  Scatterers}
\author{\IEEEauthorblockN{Ha An Le, Trinh Van Chien, \textit{Member}, \textit{IEEE},  Van Duc Nguyen, and Wan Choi, \textit{Fellow}, \textit{IEEE} \vspace{-0.8cm}}
\thanks{This work was supported by the New Faculty Startup Fund from Seoul National University. An earlier version of this paper has been submitted in part at the GLOBECOM 2023.}
\thanks{H. A. Le and W. Choi are with the  Institute of New Media and Communications and Department of Electrical and Computer Engineering, Seoul National University (SNU), Seoul 08826, (e-mail: \{25251225, wanchoi\}@snu.ac.kr) (Corresponding author: Wan Choi.)}
\thanks{T. V. Chien is with the School of Information and Communication Technology (SoICT), Hanoi University of Science and Technology (HUST), Hanoi 100000, Vietnam (email: chientv@soict.hust.edu.vn).}
\thanks{V. D. Nguyen is with the School of Electrical and Electronic Engineering (SEEE), Hanoi University of Science and Technology (HUST), Hanoi 100000, Vietnam (email: duc.nguyenvan@hust.edu.vn).}

}

\maketitle

\begin{abstract}
This paper investigates double RIS-assisted MIMO communication systems over Rician fading
channels with finite scatterers, spatial correlation, and the existence of a double-scattering link between the transceiver. First, the statistical information is driven in closed form for the aggregated channels, unveiling various influences of the system and environment on the average channel power gains. Next, we study two active and passive beamforming designs corresponding to two objectives. The first problem maximizes channel capacity by jointly optimizing the active precoding and combining matrices at the transceivers and passive beamforming at the double RISs subject to the transmitting power constraint. In order to tackle the inherently non-convex issue, we propose an efficient alternating optimization algorithm (AO) based on the alternating direction method of multipliers (ADMM). The second problem enhances communication reliability by jointly training the encoder and decoder at the transceivers and the phase shifters at the RISs. Each neural network representing a system entity in an end-to-end learning framework is proposed to minimize the symbol error rate of the detected symbols by controlling the transceiver and the RISs’ phase shifts. Numerical results verify our analysis and demonstrate the superior improvements of phase shift designs to boost system performance.

\end{abstract}
\section{Introduction}
Recent years have witnessed a significant growth of new mobile applications, as well as the number of mobile devices in a dense area, which requires the future generation wireless systems to support higher capacity, data rate, and massive connectivity \cite{WSaad2019-6GVision}. Various advanced technologies including Massive multiple-input multiple-output (MIMO), millimeter wave (mmWave) communication, or ultra-dense networks (UND) have been proposed for future wireless systems to achieve these challenging goals \cite{Andrews2014a, van2018joint}. However, those technologies are still limited by several critical issues, such as high hardware costs and increased energy consumption due to the need for more active antennas or expensive radio frequency (RF) chains operating at very high-frequency bands. To tackle the above challenges while satisfying the ever-increasing demands of future wireless systems, reconfigurable intelligent surface (RIS) was proposed as the promising solution for the sixth-generation (6G) system \cite{Ebasar2019-RISintro}.

RIS is a planar metasurface comprising a large number of passive reflecting elements that can be configured to induce the independent amplitude attenuation and/or phase shift to the incident signal, thereby collaboratively changing the wireless channels between transmitters and receivers. Therefore, RIS is a promising technology capable of modifying wireless propagation in the desired manner. This feature makes RIS distinguish from existing transmission techniques that can only optimize the wireless system over random channel conditions. Furthermore, due to the properties of being low profile and power consumption, the RIS can be manufactured in a compact size with lightweight, resulting in easy deployment.
\vspace*{-0.5cm}
\subsection{Prior Works}
Due to the promising potential, many researchers have recently been attracted by RIS technology in industry and academia. Many works combined the RIS technology with the existing systems, such as multi-cell networks \cite{CPan2020-RISmulticell},  millimeter waves \cite{PWang2020-RISmmWave}, cell-free massive MIMO \cite{9665300}, and so on.
Regarding the RIS design for MIMO systems, a joint design of the transmit beamforming and the  reflecting elements was proposed in \cite{QWu2019-RIS} that solves the total transmitted power minimization via the alternating optimization method. 
%The authors in \cite{SZhang2020-RISCapacity} studied a capacity maximization problem for RIS-assisted point-to-point MIMO systems and designed an alternating optimization algorithm to jointly control the RIS reflecting coefficients and the transmit covariance matrix. 
In addition, the RIS phase shifts were designed in \cite{JYe2020-RisSER} to enhance communication reliability by minimizing the symbol error rate (SER) of the detected signals. Furthermore, the deployment of low-cost power-efficient RISs was considered to cooperatively improve the performance of wireless networks \cite{YGao2021-DistributedRIS,Zheng2021-DoubleRIS,ZYang2022-DistributedRIS}. The authors in \cite{Zheng2021-DoubleRIS} studied a double-RIS assisted multi-user multiple-input single-output (MU-MISO) system under cooperative passive beamforming effects. It was analytically proven that a double-RIS cooperative system can perform better than conventional single-RIS systems. Then, a joint design for both the transmit beamforming and distributed phase shift control was proposed in \cite{ZYang2022-DistributedRIS} to maximize the energy efficiency of the MISO network. Instead of controlling all the available RISs in a network, the authors in \cite{ZYang2022-DistributedRIS} proposed a greedy 
 searching method to obtain the optimal RISs on-off status that leads to an online RIS controlling mechanism in an iterative manner. However, there is only a limited work \cite{Han2022-DoubleRISMIMO} on the design of double RIS-assisted MIMO systems due to the more challenging system setup. In \cite{Han2022-DoubleRISMIMO}, a design of double-RIS-aided MIMO systems was proposed to maximize the system capacity under the line-of-sight channel model. By exploiting LoS channel properties, a desirable performance could be achieved. Nevertheless, in reality, there exist both LoS and non-LoS channel that makes the design no longer applicable. Therefore, a design for double RIS-assisted MIMO systems under a general channel model is necessary. 

Although the design of RIS-assisted MIMO systems has been well studied in the literature, conventional designs suffer from high computational complexity which increases exponentially with the number of RIS reflecting elements. On the other hand, in recent years, model-free machine learning (ML) has stood out as a remarkably promising technology to address the mathematically intractable non-linear non-convex problems and high computational cost issues \cite{IGoodfellow2016-DeepLearning,Mnih2015-RL}. A great deal of effort has been made to apply ML models for the design and optimization of wireless communication systems \cite{CJiang2017-MLWireless}. Particularly, deep-learning (DL) models have been applied to optimize the beamforming matrix for MIMO systems by a mapping from the channel information to the optimal precoding matrix \cite{HHoang2019-DLMIMO},\cite{XLI2019-DLMIMO}.  Regarding the applications for the RIS design, a deep quantization neural network (DQNN) was proposed in \cite{WXU2022-RISDQNN} to jointly optimize the precoding matrix at the base station and the RIS discrete phase shifts considering the imperfect channel information. Besides, a convolutional neural network (CNN) was used in \cite{HSONG2021-RISUnsuperved} to handle a similar issue as \cite{WXU2022-RISDQNN}, but with an unsupervised fashion in order to maximize the sum rate of multi-user MIMO systems. 
%Since obtaining the channel state information (CSI) in RIS-assisted systems is particularly difficult,  the channel model-free based machine learning design has gained much attention from researchers. 
Moreover, in \cite{CHuang2020-RISDRL}, a deep reinforcement learning model was proposed to jointly learn and predict the optimal precoding matrix and RIS phase shift coefficients through a trial-and-error process that maximizes the sum rate of multiple users in the presence of a RIS. 
%After learning the properties of the environment, the proposed deep reinforcement learning model yielded a solution without any explicit knowledge of the channel model. %Therefore, ML is a very promising candidate for RIS-assisted MIMO system design. 
Regarding to communication reliability enhancement, the authors in \cite{Jiang2022-RISAutoencoder} proposed an end-to-end framework for RIS-assited MIMO
systems to improve data detection performance at the receiver. However, the work in \cite{Jiang2022-RISAutoencoder} is limited since the proposed scheme are trained based on one deterministic channel realization. Although ML models have been widely applied to control RISs' phases, we have noticed that machine learning-based designs for distributed RIS-assisted MIMO systems have not been studied due to several critical reasons. First, a conventional machine learning model requires channel information to control wireless network's entities. However, in distributed RIS systems, gathering channel information of all links is very difficult and impractical. Second, in distributed RIS systems, the number of parameters to be optimized is significant. Therefore, simple machine learning models may fail to learn a desirable solution for distributed RIS systems.
\vspace*{-0.5cm}
\subsection{Contributions}
\begin{figure} [t]
    \centering
    \includegraphics[trim=0cm 0cm 0cm 0cm, clip=true, width=3.2in]{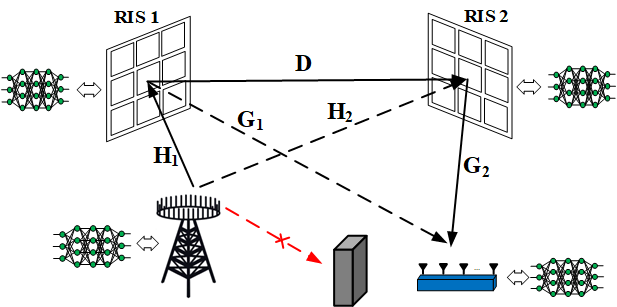}
    \caption{The considered double-RIS system model where each RIS is placed at the transmitter's and receiver's side to assist the transmission quality.}
    \label{fig:SystemModel}
    \vspace{-0.25cm}
\end{figure}

Motivated by the aforementioned issues, in the paper, we propose effective solutions for RIS-aided systems with different objectives. In particular, we consider a double RISs-assisted point-to-point MIMO system as shown in Fig.~\ref{fig:SystemModel}, where the distributed RISs are located near the transmitter and the receiver, respectively side-by-side to support the data transmission. As the channel capacity and data detection accuracy are aligned with the most important objectives of wireless communication systems, we study the joint design of the RISs' phase shifts, precoding matrix, and combining matrix at the transceiver to maximize the capacity as well as minimize the symbol error rate (SER) of the considered system. For the capacity maximization problem, an alternating optimization algorithm based on the ADMM method is proposed to tackle the non-convex objective function. For the SER minimization problem, we develop an end-to-end framework, where the transmitter, receiver, and RISs are replaced by deep neural networks, to find a feasible solution effectively. To deal with a large number of optimized parameters in double RIS systems, we proposed a one-dimensional convolutional neural network (1D-CNN)-based end-to-end design to reduce the complexity of the proposed framework. The data stream is then designed in a feasible shape and fed to the end-to-end model, and the neural networks are jointly trained in a supervised fashion to reduce the global loss of the system. The main contributions of the paper are listed as follows:
\begin{itemize}
    \item First, we characterize the propagation environment of the double RIS-assisted MIMO system by exploiting the double scattering channel model adapting to the double link between two RISs. The upper bound of the effective channel power gain is analyzed as the function of system parameters for a given set of  reflecting elements. We observe from the upper bound that when the line-of-sight (LoS) components are dominant, the channel power gains are scaled up faster with the increasing number of  reflecting elements. Moreover, for special scenarios regarding the non-line-of-sight (NLoS) channels, increasing number of scatterers will degrade the effective channel power gains.
    
    \item Second, we formulate the optimization problem that maximizes the capacity of the effective MIMO channels. In order to handle the non-convexity caused by the strong coupling between the optimization variables, we apply the singular value decomposition (SVD) method and ADMM scheme to jointly optimize the active and passive beamforming vectors. After finding the precoding and combining matrices by SVD, the phase shift elements are obtained by the alternating optimization scheme, where each variable is updated by the ADMM method to maximize the sum-path-gain of the effective MIMO channels. 
    \item Next, we formulate the optimization problem that enhances data detection reliability and solve it by proposing a novel end-to-end framework designed for the double RISs-assisted MIMO systems. In the proposed framework, each entity of the communication system is modeled as a 1D-CNN that is jointly trained to minimize symbol detection errors. Unlike the previous work \cite{HaAn2022-Autoencoder} where channel information of all reflecting links need to be gathered for ML models, the proposed end-to-end framework requires only cascaded channel information at the receiver in the inference phase, which releases the burden of channel feedback overhead. The end-to-end framework can then flexibly encode and decode data with any input length. 
    \item Finally, we present extensive simulation results to validate the performance superiority of the proposed designs over the other benchmarks in terms of both the channel capacity and SER. Moreover, the numerical results also validate our theoretical analysis of the influences of the channel parameters on the system performance.
\end{itemize}
The rest of this paper is organized as follows:
Section~\ref{Sec:SysModel} presents the system model and analyzes the power scaling law in the presence of the double scattering fading channels. Section~\ref{Sec:ChanCapOpt} formulates the channel capacity maximization problem and proposes an alternating optimization scheme to solve the formulated problem. Section~\ref{Sec:ComReliOpt} proposes the end-to-end framework to jointly control the transceivers and RISs' phase shifters to enhance the data detection reliability of the system. Simulation results and the discussion are presented in Section~\ref{Sec:SiRe} and finally, Section~\ref{Sec:Conclusion} provides main conclusions.

\textit{Notation}: The upper and lower bold letters are utilized to denote the matrices and vectors. The superscript $(\cdot)^T$ and $(\cdot)^H$ are the regular and Hermitian transpose. $\mathbf{I}_N$ denotes an identity matrix of size $N \times N$ and $\angle(\mathbf{x})$  is a vector where each element is the phase of the corresponding element in $\mathbf{x}$. The operator $\mathrm{diag}(x_1, \ldots, x_N)$ forms a diagonal matrix with the elements $x_i, \forall i =1, \ldots, N$ on the diagonal. For a square matrix $\mathbf{A}$,  $\mathrm{tr}(\mathbf{A})$ and $\mathbf{A}^{-1}$ denote its trace, and inverse, respectively. 
  $\| \cdot \|$ and $\| \cdot \|_F$ denote the Euclidean and Frobenius norm. The expectation of a random variable is denoted by $\mathbb{E}\{\cdot\}$, while $\mathcal{CN}(\cdot, \cdot)$ is a circularly symmetric Gaussian distribution. Finally, $\mathcal{O}(\cdot)$ is the big-$\mathcal{O}$ notation.
\section{System Model} \label{Sec:SysModel}
We consider a point-to-point MIMO system where a transmitter equipped with $N_t$ antenna transmits $N_s$ data streams to the receiver having $N_r$ antennas. The antenna arrays at the transmitter and receiver are assumed to be arranged in the form of  uniform linear array (ULA)\footnote{The transceiver antennas are active and costly. Consequently, the ULA structure is rationale for a moderate number of antennas. In contrast, if the number of passive  reflecting elements equipped at the RISs is considerable, the UPA structure is more reasonable.}.   The transmission is assisted by two RISs comprising of $K_1$ and $K_2$ passive  reflecting elements, respectively, which are arranged in the form of uniform planar array (UPA) as illustrated in Fig.~\ref{fig:SystemModel}. Their phase shift matrices are formulated as\footnote{In our paper, we focus solely on the double-reflection link that occurs between the transmitter and the receiver through RIS 1 and RIS 2. While another double-reflection link exists between the transmitter and the receiver through RIS 2 and RIS 1, it incurs a severe path loss due to its longer propagation distance, and thus we have ignored it in this paper.}
\begin{equation}
    \mathbf{\Phi}_i = \mathrm{diag}(\beta_{i,1}e^{j\theta_{i,1}},\cdots,\beta_{i,K_i}e^{j\theta_{i,K_i}}) = \mathrm{diag}(\mathbf{v}_i^\ast), \quad i = 1,2,
\end{equation}
where $\mathbf{v}_i^\ast = (\beta_{i,1}e^{j\theta_{i,1}},\cdots,\beta_{i,K_i}e^{j\theta_{i,K_i}})$; $(.)^\ast$ denotes the conjugate operation;  $\beta_{i,k}$ ($0\leq\beta_{i,k}\leq 1$) and $\theta_{i,k}$ ($-\pi \leq \theta_{i,k} \leq \pi$) denote the magnitude and the phase of the $k$-th reflecting element at the $i$-th RIS, respectively. Due to the recent advances towards the lossless meta-surfaces, we assume a unit signal reflection, i.e., $\beta_{i,k} = 1, \forall i,k$\cite{QWu2019-RIS}. Moreover, in this paper, we consider the challenging scenario where the direct link between the transmitter and the receiver is assumed to be blocked due to obstacles (e.g., indoor environment where direct channel is blocked by walls and corners). This harsh propagation scenario concentrates on the contributions of the double RISs. At the transmitter, data is first modulated using $M$-QAM to formulate the signal $\mathbf{s}$ with $\mathbb{E}\{\mathbf{s}\mathbf{s}^H\} = \mathbf{I}_{N_s}$. The signal is then precoded by a linear precoder $\mathbf{F} \in \mathbb{C}^{N_t \times N_s}$ satisfying $\|\mathbf{F}\|_F^2 = N_s$, and transmitted to the receiver  through the RISs. We define $\mathbf{H}_i\in \mathbb{C}^{K_i \times N_t}, i = 1,2$ as the channels between the transmitter and the $i$-th RIS, and $\mathbf{G}_i \in \mathbb{C}^{N_r \times K_i}, i = 1,2$  as the channels between the $i$-th RIS and the receiver, while $\mathbf{D} \in \mathbb{C}^{K_2 \times K_1}$ is the channel between the two RISs. At the receiver side, the received signal, denoted by $\mathbf{y} \in \mathbb{C}^{N_r}$, is expressed as
\begin{equation}
     \mathbf{y} = \sqrt{\frac{P}{N_s}}(\mathbf{G}_2\mathbf{\Phi}_2\mathbf{D}\mathbf{\Phi}_1\mathbf{H}_1 + \mathbf{G}_1\mathbf{\Phi}_1\mathbf{H}_1 + \mathbf{G}_2\mathbf{\Phi}_2\mathbf{H}_2)\mathbf{F}\mathbf{s} + \mathbf{n},
\end{equation}
where $P$ is the total transmitted power and $\mathbf{n} \sim \mathcal{CN}(\mathbf{0},\sigma^2\mathbf{I}_{N_r})$ denotes additive white Gaussian noise (AWGN). In this paper, we assume that all the channel information is known by both the transmitter and receiver with the help of channel reciprocity and sufficient resources dedicated to the pilot training process.\footnote{Note that channel estimation for the double-RIS cooperative system is a challenging task due to the presence of both single- and double-reflection links. Recently, an efficient channel estimation scheme has been proposed in \cite{ArdahArdah-DoubleRISCE}, which achieves high accuracy with practically low training overhead. Furthermore, for the optimization problem, the aggregated channels are sufficient, as demonstrated in Section III. For the estimation of these channels, one can refer to the proposed scheme in \cite{Zheng2021-DoubleRISCE2}}. The received signal is post-processed by a combiner matrix as
\begin{equation}
    \mathbf{z} = \mathbf{W}\mathbf{y} = \sqrt{\frac{P}{N_s}}\mathbf{W}\mathbf{O}\mathbf{F}\mathbf{s} + \mathbf{W}\mathbf{n},
\end{equation}
where $\mathbf{W} \in \mathbb{C}^{N_s\times N_r}$ denotes the combiner matrix satisfying $||\mathbf{W}||^2_F = N_s$, and
\begin{equation} \label{eq:H}
\mathbf{O} = \mathbf{G}_2\mathbf{\Phi}_2\mathbf{D}\mathbf{\Phi}_1\mathbf{H}_1 + \mathbf{G}_1\mathbf{\Phi}_1\mathbf{H}_1 + \mathbf{G}_2\mathbf{\Phi}_2\mathbf{H}_2
\end{equation}
is the aggregated channel of the MIMO system. The active beamforming matrices, i.e., the precoding matrix $\mathbf{F}$ and the combiner matrix $\mathbf{W}$, together with the passive reflecting matrices $\mathbf{\Phi}_1$ and $\mathbf{\Phi}_2$ can be optimized for a given particular utility metric and practical constraints. Different from most of the related literature, we consider the double scattering channel model in the RIS-assisted MIMO system by taking into account limited-scattering environments. Next, we propose two optimization approaches for each entity of the considered system, i.e. $\mathbf{F},\mathbf{W}, \mathbf{\Phi}_1$ and $\mathbf{\Phi}_2$,  in terms of the channel capacity as well as the signal detection performance.
\vspace{-0.2cm}
\subsection{Propagation Channel Model}
\begin{figure} [t]
    \centering
    \includegraphics[trim=0cm 0cm 0cm 0cm, clip=true, width=2.5in]{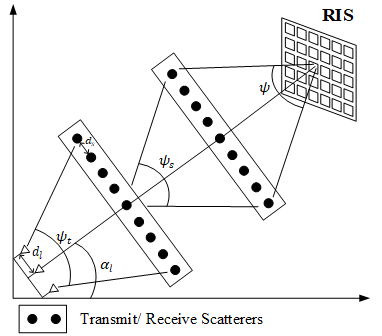}
    \caption{An example of double scattering environment between the transmitter and a RIS.}
    \label{fig:ScattererModel}
\end{figure}
%Even though the propagation channels vary over the time and frequency plane, 
This paper considers narrow-band MIMO systems where the channels are fitted into coherence blocks and are approximately static in the time domain and flat in the frequency domain. For the sake of generality, the Rician fading model is applied to all the channel links $\mathbf{A} \in \{\mathbf{H}_i,\mathbf{G}_i,\mathbf{D}\}$, $ i = 1,2$ as
\begin{align}\label{eq:channelModel}
    &\mathbf{A} = \sqrt{\chi}\left(\sqrt{\frac{\kappa}{\kappa+1}}\mathbf{\Bar{A}}+\sqrt{\frac{1}{\kappa+1}}\mathbf{\tilde{A}}\right),
\end{align}
where  $\chi \in \{\alpha_i,\beta_i,\gamma\}$ is distance-dependent large-scale path-loss factors for the corresponding link, $\kappa \in \{\epsilon_i$, $\delta_i$, $\mu\}$ is the Rician factors for each link. The matrix $\mathbf{\bar{A}}$ denotes the deterministic line-of-sight (LoS) channel, which can be expressed as the product of the UPA and ULA response vector. Specifically, the ULA response vectors for transmitter and receiver are formulated as
\begin{align} 
    \mathbf{a}_{\mathbf{H}_i}(\theta) &= [1,e^{j2\pi\frac{d_l}{\lambda}\sin(\theta)},\cdots,e^{j2\pi(N_t-1)\frac{d_l}{\lambda}\sin(\theta)}]^T\quad \label{eq:ULA-T}\\
    \mathbf{a}_{\mathbf{G}_i}(\theta) &= [1,e^{j2\pi\frac{d_l}{\lambda}\sin(\theta)},\cdots,e^{j2\pi(N_r-1)\frac{d_l}{\lambda}\sin(\theta)}]^T,  \label{eq:ULA-R}
\end{align}
where $i = 1,2,$ $\lambda$ denotes the wavelength, $d_l$ is the antenna spacing, and $\theta$ denotes the angle-of-arrival (AoA) or angle-of-departure (AoD)\cite{Hansen2009-AntennaArray}. The UPA response vectors for RIS surfaces are given as $\mathbf{a}_i(\theta,\phi) = \mathbf{a}_{vi}(\theta,\phi)\otimes \mathbf{a}_{hi}(\theta),$ where $i = 1,2,$ $\otimes$ denotes the Kronecker product. Meanwhile, $\mathbf{a}_{vi}(\theta,\phi)$ and $\mathbf{a}_{hi}(\theta)$ are the array response vectors of UPA along the two axes, which are respectively defined as follows
\begin{align}
    &\mathbf{a}_{vi}(\theta,\phi) = \notag \\
    &[1,e^{j2\pi\frac{d_v}{\lambda}\cos(\theta)\sin(\phi)},\cdots,e^{j2\pi(K_{vi}-1)\frac{d_v}{\lambda}\cos(\theta)\sin(\phi)}]^T, \label{eq:ULA-1}\\
    &\mathbf{a}_{hi}(\theta) = [1,e^{j2\pi\frac{d_h}{\lambda}\sin(\theta)},\cdots,e^{j2\pi(K_{hi}-1)\frac{d_h}{\lambda}\sin(\theta)}]^T\label{eq:ULA-2},    
\end{align}
where $K_i = K_{vi} \times K_{hi}$ is the size of the $i$-th RIS, $\theta$ and $\phi$ denote the azimuth AoA/AoD and elevation AoA/AoD, respectively, and $d_v$ and $d_h$ are the distance between two adjacent RIS reflecting elements along two axes. Exploiting the ULA and UPA response vectors defined in \eqref{eq:ULA-T} - \eqref{eq:ULA-2}, the LoS channel links are computed as follows
\begin{align} 
        &\mathbf{\bar{H}}_i=\mathbf{a}_i(\theta^A_\mathrm{Ti},\phi^A_\mathrm{Ti})\mathbf{a}_{\mathbf{H}_i}(\theta^D_\mathrm{Ti}), \label{eq:LoSv1} \\
      &\mathbf{\bar{G}}_i=\mathbf{a}_{\mathbf{G}_i}(\theta^A_\mathrm{Ri})\mathbf{a}_i(\theta^D_\mathrm{Ri},\phi^D_\mathrm{Ri}), \label{eq:LoSv2} \\
        &\mathbf{\bar{D}}=\mathbf{a}_2(\theta^A_\mathrm{R},\phi^A_\mathrm{R})\mathbf{a}_1(\theta^D_\mathrm{R},\phi^D_\mathrm{R}). \label{eq:LoSv3}
    \end{align}
In \eqref{eq:LoSv1}-\eqref{eq:LoSv3}, for the azimuth and elevation angles, i.e $\theta$ and $\phi$, the superscripts $D$ and $A$ denote the AoD and AoA, respectively. Apart from these, the subscripts denote the corresponding channel link.  

In contrast, $\mathbf{\tilde{H}}_i$, $\mathbf{\tilde{G}}_i$, $\mathbf{\tilde{D}}$ are the non-line-of-sight (NLOS) channels.\footnote{The NLoS channels have been defined by uncorrelated Rayleigh fading models in many related works with the presence of RISs. Nonetheless, Rayleigh fading models only appear in rich scattering environments. Poor scattering environments characterized by a finite number of scatterers are also of paramount importance in practice \cite{van2021uplink, Hoydis2011-DoubleScatter}.} In this paper, we consider the double scattering channel model which is a generalized version of many previous works. In particular, the NLOS channels for channel link $\mathbf{A} \in \mathbb{C}^{N_1\times N_2}$, where $N_1, N_2 \in \{N_t,N_r,K_1,K_2\}$ are the size of matrix $\mathbf{A}$ which vary with different channel links, can be generally formulated as
\begin{equation}\label{eq:doubleScatter}
    \begin{split}
        \mathbf{\tilde{A}} &= \frac{1}{\sqrt{\mathsf{SC}_{\mathbf{A}}}}\mathbf{R}_{t,\mathbf{A}}^{\frac{1}{2}}\mathbf{Q}_{\mathbf{A}}\mathbf{S}_{\mathbf{A}}^{\frac{1}{2}}\mathbf{P}_{\mathbf{A}}\mathbf{R}_{r,\mathbf{A}}^{\frac{1}{2}} ,
    \end{split} 
\end{equation}
where the subscript $\mathbf{A} \in \{\mathbf{H}_i,\mathbf{G}_i,\mathbf{D}\}$ denotes the corresponding channel link and  $\mathsf{SC}_{\mathbf{A}}$ is the number of scatterers located in the propagation channel link regarding the matrix $\mathbf{A}$. Besides, $\mathbf{R}_{t,\mathbf{A}} \in \mathbb{C}^{N_1\times N_1}$, $\mathbf{S}_{\mathbf{A}} \in \mathbb{C}^{\mathsf{SC}_{\mathbf{A}}\times \mathsf{SC}_{\mathbf{A}}}$, and $\mathbf{R}_{r,\mathbf{A}}\in \mathbb{C}^{N_2 \times N_2}$ are the transmit, scatterer, and receive correlation matrices for channel link $\mathbf{A}$, respectively.  In addition,  $\mathbf{Q}_{\mathbf{A}} \in \mathbb{C}^{N_1 \times \mathsf{SC}_{\mathbf{A}}}$, and $\mathbf{P}_{\mathbf{A}} \in \mathbb{C}^{\mathsf{SC}_{\mathbf{A}} \times N_2}$ represent the small-scale fading between the transmit side and the receive side to their scattering clusters, and $(\cdot)^{\frac{1}{2}}$ denotes the principal square root of a matrix. Without limiting the generality, we assume that the scatterers are arranged in a linear array structure as shown in Fig.~\ref{fig:ScattererModel}. Similar to \cite{Gesbert2002-OutdoorMIMO}, the correlation matrices between the transmitter and scatterers are given as follows
\begin{multline}\label{eq:Correlation}
    [\mathbf{R}_{t,\mathbf{H}_i}]_{m,n} = 
    \frac{1}{\mathsf{SC}_{\mathbf{H}_i}} \times \\ \sum_{k= 0.5\left(1-\mathsf{SC}_{\mathbf{H}_i}\right)}^{0.5\left(\mathsf{SC}_{\mathbf{H}_i}-1 \right)}\exp ( j2\pi d_t(m-n)\sin(\nu_t)),    
\end{multline}
where $[\mathbf{R}_{t,\mathbf{H}_i}]_{m,n}$ denotes the $(m,n)$-th element of the matrix $\mathbf{R}_{t,\mathbf{H}_i}$; $\psi_t$ is the angular spread of the signals; $d_t$ is the antenna spacing at the transmitter; $\mathsf{SC}_{\mathbf{H}_i}$ is the number of scatterers corresponding to channel $\mathbf{H}_i$; and $\nu_t = k\psi_t / (1-\mathsf{SC}_{\mathbf{H}_i})$. The correlation matrices between receiver and scatterers can be calculated using \eqref{eq:Correlation} with the corresponding parameters. Furthermore, the correlation matrices between the RISs and scatterers follow the spatial correlation matrices of a planar antenna array. We assume that the correlation matrix between the $i$-th RIS and the scatterers is $\mathbf{R}_i$, defined as $\mathbf{R}_i = \mathbf{R}_{vi} \otimes \mathbf{R}_{hi}$\footnote{It is shown in \cite{Emil2021-RISCorrelation} that the conventional Kronecker model causes miscalculation of the rank of the correlation matrix by a constant compared to their correlation model specifically designed for the RIS system. However, the correlation model in \cite{Emil2021-RISCorrelation} is developed under assumption of infinite number of scatterers, which is appropriate in rich scattering environments.  In poor scattering environments, the effects of a limited number of scatters must be considered. Therefore, we use the Kronecker model to characterize the correlation at the RIS in this paper.}
%Please note that a correlation model specifically designed for the RIS system is proposed in\cite{Emil2021-RISCorrelation}. The proposed correlation model demonstrates that the conventional Kronecker model is unsuitable for modeling correlation in the asymptotic region of the RIS. However, the correlation model in \cite{Emil2021-RISCorrelation} is developed under assumption of infinite number of scatterers, which is not the case in this paper. Furthermore, our study evaluates the system in the non-asymptotic region, where the number of RIS reflecting elements is not significant. Therefore, we use the Kronecker model to characterize the correlation at the RIS in our paper.}}
, where $\mathbf{R}_{vi}$ and $\mathbf{R}_{hi}$ express the correlation along the vertical and horizontal directions of the $i$-th RIS, for a particular $\mathbf{R}_i \in \{\mathbf{R}_{r,\mathbf{H}_i},\mathbf{R}_{t,\mathbf{G}_i},\mathbf{R}_{r,\mathbf{D}},\mathbf{R}_{t,\mathbf{D}}\}$. Note that along each direction,  $\mathbf{R}_{vi}$ and $\mathbf{R}_{hi}$ refer to a subset of RIS elements arranged in ULA structure, thus the correlation matrices along each axis of the RIS are given as
\begin{align}
     & [\mathbf{R}_{vi}]_{m,n} =
    \frac{1}{\mathsf{SC}_{\mathbf{A}}} \times \notag \\
    & \qquad \qquad \sum_{k=0.5 \left(1-\mathsf{SC}_{\mathbf{A}} \right)}^{0.5 \left( \mathsf{SC}_{\mathbf{A}}-1 \right)}\exp\left(j2\pi d_v(m-n)\sin\left(\frac{k\psi_i}{1-\mathsf{SC}_{\mathbf{A}}}\right)\right),\\
    & [\mathbf{R}_{hi}]_{m,n} =
    \frac{1}{\mathsf{SC}_{\mathbf{A}}}  \times \notag \\
    & \qquad \qquad \sum_{k=0.5(1-\mathsf{SC}_{\mathbf{A}})}^{0.5(\mathsf{SC}_{\mathbf{A}}-1)}\exp\left(j2\pi d_h(m-n)\sin\left(\frac{k\psi_i}{1-\mathsf{SC}_{\mathbf{A}}}\right)\right).
\end{align}
Finally, the correlation matrices between the scatterers are calculated as 
\begin{multline}
    [\mathbf{S}_{\mathbf{A}}]_{m,n} =
    \frac{1}{\mathsf{SC}_{\mathbf{A}}} \times \label{eq:Hi}\\ \sum_{k=0.5(1-\mathsf{SC}_{\mathbf{A}})}^{0.5(\mathsf{SC}_{\mathbf{A}}-1)}\exp\left(j2\pi d_s(m-n)\sin\left(\frac{k\psi_s}{1-\mathsf{SC}_{\mathbf{A}}}\right)\right),    
\end{multline}
where $d_s$ is the distance between two scatterers, and $\psi_{s}$ is the angular spread.
\begin{remark}
The double scattering channel model considered in this section is a general model where the correlation matrices are modeled corresponding to the antenna and RIS structures, i.e., ULA and UPA, and covers various channel environments from uncorrelated Rayleigh to keyhole channel. The uncorrelated Rayleigh channel is obtained by setting $\mathbf{R}_{t,\mathbf{A}} = \mathbf{I}_{N_1}, \mathbf{R}_{r,\mathbf{A}} = \mathbf{I}_{N_2}, \mathbf{S}_{\mathbf{A}} = \mathbf{I}_{\mathsf{SC}_{\mathbf{A}}}$, and letting $\mathsf{SC}_{\mathbf{A}}\rightarrow \infty$, which is the assumption of perfect scattering condition with high-rank correlation matrices. On the other hand, $\mathsf{SC}_{\mathbf{A}} = 1$ yields a keyhole channel with the worst case of rank-deficiency, i.e $\mathbf{R}_{t,\mathbf{A}}, \mathbf{R}_{r,\mathbf{A}}, \mathbf{S}_{\mathbf{A}}$ have rank $1$.
\end{remark}
\vspace{-0.3cm}
\subsection{Power Scaling Law}
In this subsection, we study the power scaling law of the propagation channels by analyzing their statistical properties.
\begin{lemma} \label{Lemma1}
 For a given set of the phase-shift coefficients, the covariance matrix of the cascaded channel of  double RIS-assisted MIMO systems is computed in a closed-form expression as follows
 \begin{align} \label{eq:ChannelSecondMoment}
     &\mathbb{E}\left[\mathbf{O}\mathbf{O}^H\right] = \alpha_1\beta_2\gamma\frac{\mu\delta_2}{(\mu+1)(\delta_2+1)}\mathbf{\bar{G}}_2\mathbf{\Phi}_2\mathbf{\bar{D}}\mathbf{X}_1\mathbf{\bar{D}}^H\mathbf{\Phi}_2^H\mathbf{\bar{G}}_2^H \notag\\
   &+\alpha_1\beta_2\gamma\frac{\mu}{(\mu+1)(\delta_2+1)}\mathrm{tr}\left(\mathbf{R}_{r,\mathbf{G}_2}^{\frac{1}{2}}\mathbf{\Phi}_2\mathbf{\bar{D}}\mathbf{X}_1\mathbf{\bar{D}}^H\mathbf{\Phi}_2^H\mathbf{R}_{r,\mathbf{G}_2}^{\frac{1}{2}}\right)\mathbf{R}_{t,\mathbf{G}_2} \notag\\
   &+\alpha_1\beta_2\gamma\frac{\delta_2}{(\mu+1)(\delta_2+1)}\mathrm{tr}\left(\mathbf{R}_{r,\mathbf{D}}^{\frac{1}{2}}\mathbf{X}_1\mathbf{R}_{r,\mathbf{D}}^{\frac{1}{2}}\right)\mathbf{\bar{G}}_2\mathbf{\Phi}_2\mathbf{R}_{t,\mathbf{D}}\mathbf{\Phi}_2^H\mathbf{\bar{G}}_2^H \notag\\
   &+ \alpha_1\beta_2\gamma\frac{1}{(\mu+1)(\delta_2+1)}\mathrm{tr}\left(\mathbf{R}_{r,\mathbf{D}}^{\frac{1}{2}}\mathbf{X}_1\mathbf{R}_{r,\mathbf{D}}^{\frac{1}{2}}\right) \times \notag\\
    &\quad \quad \mathrm{tr}\left(\mathbf{R}_{r,\mathbf{G}_2}^{\frac{1}{2}}\mathbf{\Phi}_2\mathbf{R}_{t,\mathbf{D}}\mathbf{\Phi}_2^H\mathbf{R}_{r,\mathbf{G}_2}^{\frac{1}{2}}\right)\mathbf{R}_{t,\mathbf{G}_2} \notag\\
    &+2\sqrt{\alpha_1^2\beta_1\beta_2\gamma}\sqrt{\frac{\delta_1 \delta_2 \mu}{(\delta_1+1)(\delta_2+1)(\mu+1)}}  \mathrm{Re}\big\{\mathbf{\bar{G}}_2\mathbf{\Phi}_2\mathbf{\bar{D}}\mathbf{X}_1\mathbf{\bar{G}}^H_1\big\} \notag\\
    &+2\sqrt{\alpha_1\alpha_2\beta_2^2\gamma}\sqrt{\frac{\epsilon_1 \epsilon_2 \mu}{(\epsilon_1+1)(\epsilon_2+1)(\mu+1)}} \times \notag\\
    &\quad \quad \mathrm{Re}\big\{\mathbf{\bar{G}}_2\mathbf{\Phi}_2\mathbf{\bar{D}}\mathbf{\Phi}_1\mathbf{\bar{H}}_1\mathbf{\bar{H}}^H_2\mathbf{\Phi}_2\mathbf{\bar{G}}^H_2\big\} \notag\\
    &+\alpha_1\beta_1\left(\frac{\delta_1}{\delta_1+1}\mathbf{\bar{G}}_1\mathbf{X}_1\mathbf{\bar{G}}^H_1+\frac{1}{\delta_1+1}\mathrm{tr}\left(\mathbf{R}_{r,\mathbf{G}_1}^{\frac{1}{2}}\mathbf{X}_1\mathbf{R}_{r,\mathbf{G}_1}^{\frac{1}{2}}\right)\mathbf{R}_{t,\mathbf{G}_1}\right) \notag\\
    &+2\sqrt{\alpha_1\alpha_2\beta_1\beta_2}\sqrt{\frac{\epsilon_1 \epsilon_2 \delta_1 \delta_2}{(\epsilon_1+1)(\epsilon_2+1)(\delta_1+1)(\delta_2+1)}} \times \notag\\
    &\quad \quad  \mathrm{Re}\big\{\mathbf{\bar{G}}_1\mathbf{\Phi}_1\mathbf{\bar{H}}_1\mathbf{\bar{H}}^H_2\mathbf{\Phi}_2^H\mathbf{\bar{G}}^H_2\big\}\notag\\
    &+ \alpha_2\beta_2 \left(\frac{\delta_2}{\delta_2+1}\mathbf{\bar{G}}_2\mathbf{X}_2\mathbf{\bar{G}}^H_2+\frac{1}{\delta_2+1}\mathrm{tr}\left(\mathbf{R}_{r,\mathbf{G}_2}^{\frac{1}{2}}\mathbf{X}_2\mathbf{R}_{r,\mathbf{G}_2}^{\frac{1}{2}}\right)\mathbf{R}_{t,\mathbf{G}_2}\right),
 \end{align}
where $\mathbf{X}_1 = \mathbf{\Phi}_1(\frac{\epsilon_1}{\epsilon_1+1}\mathbf{\bar{H}}_1\mathbf{\bar{H}}_1^H+\frac{1}{\epsilon_1+1}N_t\mathbf{R}_{t,\mathbf{H}_1})\mathbf{\Phi}_1^H$, and $\mathbf{X}_2 = \mathbf{\Phi}_2(\frac{\epsilon_2}{\epsilon_2+1}\mathbf{\bar{H}}_2\mathbf{\bar{H}}_2^H+\frac{1}{\epsilon_2+1}N_t\mathbf{R}_{t,\mathbf{H}_2})\mathbf{\Phi}_2^H$. \\

\begin{IEEEproof}
Please refer to Appendix~\ref{Appendix:Lemma1}.
\end{IEEEproof}
\end{lemma}
Lemma \ref{Lemma1} allows us to further investigate the channel power gain metric of the considered system. By using \cite[Lemma B.7]{Emil2017-MMIMOBook} and cyclic property of trace, we observe the power scaling law for the double RIS-assisted MIMO system under double scattering channel model as
\begin{align} \label{eq:PowerLaw}
        &\mathbb{E}\left\{\mathrm{tr}\big(\mathbf{O}\mathbf{O}^H\big)\right\} \leq \alpha_1\beta_2\gamma\frac{\mu\delta_2}{(\mu+1)(\delta_2+1)}N_tN_rK_1^2K_2^2 \notag \\
        &+\alpha_1\beta_2\gamma\frac{\mu}{(\mu+1)(\delta_2+1)}N_tN_rK_1K_2K_1||\mathbf{R}_{r,\mathbf{G}_2 }||_2 \notag\\
        &+\alpha_1\beta_2\gamma\frac{\delta_2}{(\mu+1)(\delta_2+1)}N_tN_rK_1K_2K_2||\mathbf{R}_{r,\mathbf{D}}||_2 \notag \\
        &+\alpha_1\beta_2\gamma\frac{1}{(\mu+1)(\delta_2+1)}N_tN_rK_1K_2||\mathbf{R}_{r,\mathbf{G}_2}||_2||\mathbf{R}_{r,\mathbf{D}}||_2\notag\\
        &+2\sqrt{\alpha_1^2\beta_1\beta_2\gamma}\sqrt{\frac{\delta_1 \delta_2 \mu}{(\delta_1+1)(\delta_2+1)(\mu+1)}}N_tK_1||\mathbf{\bar{G}}_1^H\mathbf{\bar{G}}_2\mathbf{\Phi}_2\mathbf{\bar{D}}||_2\notag\\
        &+2\sqrt{\alpha_1\alpha_2\beta_2^2\gamma}\sqrt{\frac{\epsilon_1 \epsilon_2 \mu}{(\epsilon_1+1)(\epsilon_2+1)(\mu+1)}}N_rK_2||\mathbf{\bar{D}}\mathbf{\Phi}_1\mathbf{\bar{H}}_1\mathbf{\bar{H}}_2^H||_2\notag\\
        &+\alpha_1\beta_1\left(\frac{\delta_1}{\delta_1+1}N_tN_rK_1^2+\frac{1}{\delta_1+1}N_tN_rK_1||\mathbf{R}_{r,\mathbf{G}_1}||_2\right)\notag\\
        &+2\sqrt{\alpha_1\alpha_2\beta_1\beta_2}\sqrt{\frac{\epsilon_1 \epsilon_2 \delta_1 \delta_2}{(\epsilon_1+1)(\epsilon_2+1)(\delta_1+1)(\delta_2+1)}} \times \notag \\
        &\quad \quad \mathrm{Re}\big\{\mathrm{tr}(\mathbf{\bar{G}}_1\mathbf{\Phi}_1\mathbf{\bar{H}}_1\mathbf{\bar{H}}_2^H\mathbf{\Phi}_2^H\mathbf{\bar{G}}_2^H)\big\} \notag\\
        &+\alpha_2\beta_2\left(\frac{\delta_2}{\delta_2+1}N_tN_rK_2^2+\frac{1}{\delta_2+1}N_tN_rK_2||\mathbf{R}_{r,\mathbf{G}_2}||_2\right).
\end{align}   

The result in \eqref{eq:PowerLaw} reveals that when LoS links are dominant, the sum-path-gain of cascaded channel scales like $\mathcal{O}(N_tN_rK_1^2K_2^2)$. 
Moreover, for the case of NLoS channel model, i.e. $\epsilon_i = \delta_i = \mu = 0$, the total gain of cascaded channel is bounded as
\begin{equation}\label{eq:NLoSPowerLaw}
    \begin{split}
       &\mathbb{E}\left\{\mathrm{tr}\big(\mathbf{O}\mathbf{O}^H\big)\right\} \leq \alpha_1\beta_2\gamma N_tN_rK_1K_2||\mathbf{R}_{r,\mathbf{G}_2}||_2||\mathbf{R}_{r,\mathbf{D}}||_2\\
       &+ \alpha_1\beta_1N_tN_rK_1||\mathbf{R}_{r,\mathbf{G}_1}||_2 + \alpha_2\beta_2N_tN_rK_2||\mathbf{R}_{r,\mathbf{G}_2}||_2\\
       &\stackrel[(a)]{\mathsf{SC}_A \rightarrow \infty}{\longrightarrow}\alpha_1\beta_2\gamma N_tN_rK_1K_2 + \alpha_1\beta_1N_tN_rK_1 + \alpha_2\beta_2N_tN_rK_2,
    \end{split}
\end{equation}
where $(a)$ is due to the fact that correlation matrices will converge to identity matrices when the number of scatterers becomes very large \cite{Gesbert2002-OutdoorMIMO}. The result in \eqref{eq:NLoSPowerLaw} indicates that when the NLoS links are dominant, the sum-path-gain of the effective channel scales as $\mathcal{O}(N_tN_rK_1K_2)$. In addition, increasing the number of scatterers will result in a decrease of channel energy. 
\section{Channel Capacity Optimization} \label{Sec:ChanCapOpt}
The channel capacity maximization of the RIS-assisted MIMO system, which we aim to solve, is formulated as follows

\begin{align}\label{eq:P1}
\mathrm{(P1)}:&\underset{\mathbf{F},\mathbf{W},\mathbf{v}_1,\mathbf{v}_2}{\mathrm{maximize}} 
&&\log_2\det\left|\mathbf{I}_{N_r}+\frac{P}{\sigma^2N_s}\mathbf{W}\mathbf{O}\mathbf{F}\mathbf{F}^H\mathbf{O}^H\mathbf{W}^H\right| \\
     &\mbox{subject to}&&||\mathbf{F}||_F^2=N_s,\mathbf{\Phi}_i=\mathrm{diag}(\mathbf{v}_i^\ast),\\
     & &&|\mathbf{v}_i(j)| = 1, i = 1,2, j = 1,2,\cdots,K_i.
\end{align}
It is easy to see that $\mathrm{(P1)}$ is a non-convex optimization problem due to the non-convex objective function as well as the unit-modular constraint. Moreover, the RISs' phase shifts $\mathbf{\Phi}_1$ and $\mathbf{\Phi}_2$ are coupled with the precoding matrix $\mathbf{F}$ and the combining matrix $\mathbf{W}$ which makes $\mathrm{(P1)}$ even more difficult to solve. A straightforward solution would be applying machine learning techniques to solve $\mathrm{(P1)}$ in a supervised-learning fashion. However, supervised learning requires true labels to train neural network models, which is difficult to obtain since an optimal mathematical solution for $\mathrm{(P1)}$ is not available in the literature. On the other hand, while one can apply an unsupervised learning model to solve $\mathrm{(P1)}$, the convergence point can be easily stuck at the local optima. 
In this section, in order to solve $(P1)$, we first propose a sum-path-gain criterion inspired by \cite{Bouy2020-RISsumpath} to design the phase-shift vectors in \eqref{eq:P1}, and then optimize the precoder $\mathbf{F}$ and the combiner $\mathbf{W}$ via singular value decomposition (SVD) and water-filling power allocation framework\footnote{Even though the sum-path-gain approach and ADMM framework have been utilized in \cite{Bouy2020-RISsumpath} for RIS phase shift optimization, the proposed framework  is limited for single RIS-aided systems. However, in double-RIS-assisted systems, the phase shift of the dual RISs are coupled due to the cooperative reflection link, making the problem significantly more complex. Therefore, we aim to tackle this challenging problem by leveraging the AO process.}.
\vspace{-0.3cm}
\subsection{Joint Precoder and Combiner Matrices Optimization}
In order to maximize the capacity of the MIMO channel, we formulate the SVD of 
$\mathbf{O}$ as $\mathbf{O} = \mathbf{U} \mathbf{\Lambda} \mathbf{V}^H$, where $\mathbf{U} \in \mathbb{C}^{N_r\times N_t}$, $\mathbf{V} \in \mathbb{C}^{N_t \times N_t}$ satisfying 
$\mathbf{U}^H\mathbf{U} = \mathbf{I}_{N_t}$ and $\mathbf{V}^H\mathbf{V} = \mathbf{I}_{N_t}$. In addition, $\mathbf{\Lambda} = \mathrm{diag}(\lambda_1,\lambda_2,\cdots,\lambda_{N_t})$ contains the singular values $\lambda_m, \forall m=1,2,\cdots,N_t$, with $\lambda_1\geq \lambda_2 \geq,\cdots,\geq \lambda_{N_t}$. The optimal precoding matrix is given by $\mathbf{F}_\ast = [\mathbf{V}]_{1:N_s}\mathbf{\Gamma}^{\frac{1}{2}}_{\ast}$, where $\mathbf{\Gamma}_{\ast}  \triangleq \mathrm{diag}([p_1^\ast,\cdots,p^\ast_{N_s}])$ is power allocation matrix, and $p_i^\ast$ is the optimal power allocated to the $i$-th data stream satisfying $\sum_{i=1}^{N_s}p_i=N_s$. Similarly, the combiner matrix is designed as $\mathbf{W}_{\ast} = [\mathbf{U}]_{1:N_s}^H.$
Substitute the SVD-based solutions into \eqref{eq:P1}, the objective function can be re-written as $\sum_{i=1}^{N_s}\log_2\left(1+\frac{Pp_i\lambda_i^2}{\sigma^2N_s}\right)$. Therefore, (P1)  is equivalent to
\begin{align}
\mathrm{(P2)}:&\underset{\mathbf{v}_1,\mathbf{v}_2,p_i}{\mathrm{maximize}}&&\sum_{i=1}^{N_s}\log_2\left(1+\frac{Pp_i\lambda_i^2}{\sigma^2N_s}\right) \\
    &\mbox{subject to} && \sum_{i=1}^{N_s}p_i=N_s, \\
    & &&|\mathbf{v}_i(j)| = 1, i = 1,2, j = 1,2,\cdots,K_i.
\end{align}
In $\mathrm{(P2)}$, the optimal power allocations can be obtained by the water-filling procedure with the given $\mathbf{v}_1$ and $\mathbf{v}_2$. Therefore, in the following subsections, we focus on the optimization of the phase-shift vectors, i.e., $\mathbf{v}_1$ and $\mathbf{v}_2$. 

Unfortunately, it is still intractable to optimize $\mathrm{(P2)}$ with the subject to $\mathbf{v}_1$ and $\mathbf{v}_2$, due to the implicit relationship between the objective function and $\mathbf{v}_i$, as well as the non-convex unit-modulus constraint. In order to efficiently solve $\mathrm{(P2)}$, we follow the same methodology as \cite[Proposition 1]{Bouy2020-RISsumpath}. Specifically, the solution for $\mathrm{(P2)}$ can be found by solving the following problem.
\begin{align} \label{eq:P3}
     \mathrm{(P3)}:&\underset{\mathbf{v}_1,\mathbf{v}_2}{\mathrm{maximize}} &&\mathrm{tr}\Big((\mathbf{G}_2\mathbf{\Phi}_2\mathbf{D}\mathbf{\Phi}_1\mathbf{H}_1 + \mathbf{G}_1\mathbf{\Phi}_1\mathbf{H}_1 + \mathbf{G}_2\mathbf{\Phi}_2\mathbf{H}_2)  \notag \\
     & &&(\mathbf{G}_2\mathbf{\Phi}_2\mathbf{D}\mathbf{\Phi}_1\mathbf{H}_1 + \mathbf{G}_1\mathbf{\Phi}_1\mathbf{H}_1 + \mathbf{G}_2\mathbf{\Phi}_2\mathbf{H}_2)^H \Big) \\
     &\mbox{subject to} &&\mathbf{\Phi}_1 = \mathrm{diag}(\mathbf{v}_1^\ast),\mathbf{\Phi}_2 = \mathrm{diag}(\mathbf{v}_2^\ast),\\
     & &&|\mathbf{v}_i(j)| = 1, i = 1,2, j = 1,2,\cdots,K_i.
\end{align}

The problem is, however, still non-convex due to the modulo constraints as well as the coupled $\mathbf{v}_1$ and $\mathbf{v}_2$. To solve $\mathrm{(P3)}$, we propose an AO-based framework for designing the sub-optimal cooperative reflecting beamforming from the two RISs.
\vspace{-0.3cm}
\subsection{AO Framework for Double RISs Phase-shift Design}
In this subsection, an AO-based algorithm is proposed in order to solve $\mathrm{(P3)}$. Specifically, the phase-shift vectors of the two RISs are optimized in an iterative manner, until convergence is obtained.
First, for a fixed $\mathbf{v}_2$, $\mathrm{(P3)}$ is equivalent to
\begin{align} \label{eq:phi1}
    \mathrm{(P3.1)}:&\underset{\mathbf{v}_1}{\mathrm{maximize}}&&\mathrm{tr}\Big((\mathbf{M}\mathbf{\Phi}_1\mathbf{H}_1+\mathbf{A})(\mathbf{M}\mathbf{\Phi}_1\mathbf{H}_1+\mathbf{A})^H\Big)\\
    &\mbox{subject to} &&\mathbf{\Phi}_1 = \mathrm{diag}(\mathbf{v}_1^\ast),|\mathbf{v}_1(j)| = 1,  j = 1,\cdots,K_i,\\
    & &&\mathbf{M} = \mathbf{G}_2\mathbf{\Phi}_2\mathbf{D} + \mathbf{G}_1, \mathbf{A} = \mathbf{G}_2\mathbf{\Phi}_2\mathbf{H}_2. 
\end{align}
The objective function can be re-written as $\sum_{i=1}^{N_r}[\mathbf{\Pi}(i,i)+\mathbf{\Upsilon}(i,i)+\mathbf{\Upsilon}(i,i)^H]+C$, where $\mathbf{\Pi} = \mathbf{M}\mathbf{\Phi}_1\mathbf{H}_1\mathbf{H}_1^H\mathbf{\Phi}_1^H\mathbf{M}^H$, $\Upsilon = \mathbf{M}\mathbf{\Phi}_1\mathbf{H}_1\mathbf{B}^H$, and $C = \mathrm{tr}(\mathbf{B}\mathbf{B}^H)$. Define $\mathbf{M} = [\mathbf{m}_1^H;\mathbf{m}_2^H;\cdots;\mathbf{m}_{N_r}^H]$, $\mathbf{H}_1\mathbf{B}^H = [\mathbf{k}_1,\mathbf{k}_2,\cdots,\mathbf{k}_{N_r}]$, then we have
\begin{equation}
    \begin{split}
        &\mathbf{\Pi}(i,i) = \mathbf{m}_i^H\mathrm{diag}(\mathbf{v}_1^\ast)\mathbf{H}_1\mathbf{H}_1^H\mathrm{diag}(\mathbf{v})\mathbf{m}_i,\\
        & \mathbf{\Upsilon}(i,i) = \mathbf{m}_i^H\mathrm{diag}(\mathbf{v}_1^\ast)\mathbf{k}_i.
    \end{split}
\end{equation}
Since $\mathbf{n}_i^H\mathrm{diag}(\mathbf{v}_1^\ast) = \mathbf{v}^H\mathrm{diag}(\mathbf{n_i})^H$, we introduce a matrix and a vector as
\begin{equation}
\hspace{-0.3cm}
    \mathbf{T} = \begin{pmatrix}
    -\sum\limits_{i=1}^{N_r}\mathrm{diag}(\mathbf{m}_i^H)\mathbf{H}_1\mathbf{H}_1^H\mathrm{diag}(\mathbf{m}_i) & -\sum\limits_{i=1}^{N_r}\mathrm{diag}(\mathbf{m}_i)\mathbf{k}_i\\
    -\sum_{i=1}^{N_r}\mathbf{k}_i^H\mathrm{diag}(\mathbf{m}_i) & 0
    \end{pmatrix},
\end{equation}
and $\mathbf{p}=[t\mathbf{v}_1^H, t]^H$, respectively, where $t\in \mathbb{C}$ is an auxiliary variable satisfying $|t| = 1$. Thus, $\mathrm{(P3.1)}$ can be equivalently rewritten as
\begin{align}
     &\underset{\mathbf{p}}{\mathrm{minimize}} &&\frac{1}{2}\mathbf{p}^H\mathbf{T}\mathbf{p} \label{eq:P31equi}\\   
     &\mbox{subject to} &&|\mathbf{p}(i)| = 1, i = 1,2,\cdots,K_1+1.
\end{align}

Moreover, by denoting $\hat{\mathbf{T}} = \mathbf{T}-\lambda_{min}(\mathbf{T})\mathbf{I}_{K_1+1} \geq \mathbf{0}$, 
where $\lambda_{min}(\mathbf{T})$ is the minimum eigenvalue of $\mathbf{T}$, $\mathrm{(P3.1)}$ is equivalent to 
\begin{align}
        &\underset{\mathbf{p},\mathbf{u}}{\mathrm{minimize}}&&\frac{1}{2}\mathbf{p}^H\hat{\mathbf{T}}\mathbf{p} \\
        &\mbox{subject to} &&|\mathbf{u}(i)| = 1, \quad i = 1,\cdots,K_1+1,\\
        & &&\mathbf{u} = \mathbf{p},
\end{align}
where $\mathbf{u}$ is a slack variable induced to fit the problem into the ADMM framework. Following the similar procedure in \cite{Qiang2019-ADMM}, we can apply the ADMM framework to solve the problem. Specifically, the augmented Lagrangian function is given by 
\begin{equation}
    \mathcal{L}(\mathbf{u},\mathbf{p},\pmb{\nu}) = \frac{1}{2}\mathbf{p}^H\hat{\mathbf{T}}\mathbf{p} + \mathrm{Re}\{\pmb{\nu}^H(\mathbf{u}-\mathbf{p})\}+\frac{\rho}{2}||\mathbf{u}-\mathbf{p}||^2,
\end{equation}
where $\pmb{\nu}\in \mathbb{C}^{K_1+1}$ is the Lagrange multiplier associated with the second equality constraint, and $\rho$ is the positive penalty parameter. Define ($\mathbf{u}^0,\mathbf{p}^0,\pmb{\nu}^0)$ as the initial primal-dual variables. The ADMM repeatedly performs the following updates:
\begin{equation}
    \begin{cases}
    \mathbf{u}^{k+1} &= \arg \min_{\mathbf{u}} \mathcal{L}(\mathbf{u},\mathbf{p}^k,\pmb{\nu}^k)\\
    \mathbf{p}^{k+1} &= \arg\min_{\mathbf{p}} \mathcal{L}(\mathbf{u}^k,\mathbf{p},\pmb{\nu}^k)\\
    \pmb{\nu}^{k+1} &= \pmb{\nu}^k + \rho(\mathbf{u}^{k+1}-\mathbf{p}^{k+1}).
    \end{cases}
\end{equation}
The solutions for the above problems can be computed as \cite{Qiang2019-ADMM}
\begin{align}
    \mathbf{u}^{k+1} &= \angle (\mathbf{p}^k-\rho^{-1}\pmb{\nu}^k) \label{eq:updateU},\\
    \mathbf{p}^{k+1} &= (\rho\mathbf{I} + \hat{\mathbf{T}})^{-1}(\rho\mathbf{u}^{k+1}+\pmb{\nu}^k) \label{eq:updateP},\\
    \pmb{\nu}^{k+1} &= \hat{\mathbf{T}}\mathbf{p}^{k+1} \label{eq:updateNu}.
\end{align}
The above steps are repeated with the increase of $k$ until convergence is achieved. It is notable that the convergence can be guaranteed with a proper initialization. Specifically, if the penalty parameter satisfies  $\rho > \sqrt{2}\lambda_{max}(\hat{\mathbf{T}})$ \cite[Proposition 1]{Qiang2019-ADMM}, where $\lambda_{max}(.)$ denotes the maximum eigenvalue of a matrix, the solution point generated by \eqref{eq:updateU}-\eqref{eq:updateNu} is a Karush-Kuhn-Tucker (KKT) point of \eqref{eq:P31equi}. The optimal phase-shift vector for RIS 1 is then computed as $\mathbf{v}_1 = \big\{\mathbf{p}^k \big[\mathbf{p}^k(K_1+1)\big]^{-1}\big\}_{1:K_1}$.

Next, we optimize the phase-shift vector $\mathbf{v}_2$ of RIS 2 with $\mathbf{v}_1$ being fixed. Problem $\mathrm{(P3)}$ is now equivalent to
\begin{align} \label{eq:phi2}
    \mathrm{(P3.2)}:&\underset{\mathbf{v}_2}{\mathrm{maximize}} &&\mathrm{tr}\Big((\mathbf{G}_2\mathbf{\Phi}_2\mathbf{N}+\mathbf{B})(\mathbf{G}_2\mathbf{\Phi}_2\mathbf{N}+\mathbf{B})^H\Big)\\
    &\mbox{subject to} &&\mathbf{\Phi}_1 = \mathrm{diag}(\mathbf{v}_2^\ast),|\mathbf{v}_2(j)| = 1,  j = 1,2,\cdots,K_i,\\
    & &&\mathbf{N} = \mathbf{G}\mathbf{\Phi}_1\mathbf{H}_1 + \mathbf{H}_2, \mathbf{B} = \mathbf{G}_1\mathbf{\Phi}_1\mathbf{H}_1. 
\end{align}
Problem $\mathrm{(P3.2)}$ can be similarly solved following the same transformation and method for the problem $\mathrm{(P3.1)}$. Finally, the AO-based algorithm is performed by iteratively solving $\mathrm{(P3.1)}$ and $\mathrm{(P3.2)}$ until the convergence is achieved.It is worth noting that this convergence is guaranteed by the following two key factors. First, during each iteration, both $\mathrm{(P3.1)}$ and $\mathrm{(P3.2)}$ increase the objective of $\mathrm{(P3)}$, causing it to become non-increasing over iterations. Second, the objective of $\mathrm{(P3)}$ is indeed upper-bounded due to the modulo constraint which prevents in from increasing indefinitely.

After the convergence, with the given $\mathbf{v}_1$ and $\mathbf{v}_2$,  the precoding and combining matrices can be obtained by the SVD procedure as mentioned in the previous subsection. Moreover, the power allocation vector can be found by applying the water-filling scheme \cite{Stephen2009-convexOpt} as $p_i^{\ast} = \left(\frac{1}{\alpha \ln 2}-\frac{\sigma^2N_s}{P\lambda_i^2}\right)^+, \quad i = 1,\cdots, N_s$, where $(.)^+ \triangleq \max\{.,0\}$, and $\alpha$ is a constant which ensures $\sum_{i=1}^{N_s}p_i^{\ast}=N_s$. The overall algorithm to solve $\mathrm{(P1)}$ is summarized in Algorithm 1. 

\begin{algorithm}[t]
\caption{AO-based Algorithm for Solving problem$\mathrm{(P1)}$}\label{alg:cap}
\begin{algorithmic}[1]
%\State \textbf{Given:}  $P,\sigma^2,\mathbf{H}_i,\mathbf{G}_i,\mathbf{D}, i=1,2.$ 
\State Initialization: $\mathbf{v}_1 := \mathbf{v}_1^{(0)}$, $\mathbf{v}_2 := \mathbf{v}_2^{(0)}$ and the iteration index $i=0$.
\State \textbf{repeat}
\State \quad Solve problem $\mathrm{(P3.1)}$ for the given $\mathbf{v}_2^{(i)}$ via ADMM framework. Denote the solution after the convergence as $\mathbf{v}_1^{(i+1)}$.
\State \quad Solve problem $\mathrm{(P3.2)}$ for the given $\mathbf{v}_1^{(i+1)}$ via ADMM framework. Denote the solution after the convergence as $\mathbf{v}_2^{(i+1)}$.
\State Update $i:=i+1$.
\State \textbf{until} The fractional increase of the objective function in \eqref{eq:P3} is less than a threshold $\epsilon > 0$.
\State Compute the optimal precoding and combining matrices $\mathbf{F}$ and $\mathbf{W}$ via SVD and power-allocation procedure.
\end{algorithmic}
\end{algorithm}

\subsection{Computational Complexity}
The computational complexity of the proposed algorithm mainly comes from phase-shift optimization. For problem $\mathrm{(P3.1)}$, we analyze the computational complexity as follows. The update of $\mathbf{u}$ in \eqref{eq:updateU} requires $\mathcal{O}(K_1+1)$ multiplication operations, while updating $\mathbf{p}$ in \eqref{eq:updateP} requires $\mathcal{O}((K_1+1)^3+(K_1+1)^2)$ multiplication operators, and the update of $\pmb{\nu}$ in \eqref{eq:updateNu} requires $\mathcal{O}(K_1+1)$ multiplication operations. It is worth noting that the calculation of $(\rho\mathbf{I} + \hat{\mathbf{T}})^{-1}$ can be performed once and used throughout the whole procedure. Therefore, the computational complexity of solving $\mathrm{P(3.1)}$ is in order of $\mathcal{O}(K_1^3+T_1K_1^2)$, where $T_1$ is the number of ADMM iterations. Similarly, the computational complexity of solving $\mathrm{P(3.2)}$ is in the order of $\mathcal{O}(K_2^3+T_2K_2^2)$, where $T_2$ is the number of iterations for solving $\mathrm{P(3.2)}$. In addition, performing SVD-based computation requires the complexity of $\mathcal{O}(N_tN_r.\mathrm{min}(N_t,N_r))$.  Overall, the computational complexity of Algorithm 1 can be evaluated as $\mathcal{O}(T(K_1^3+T_1K_1^2+K_2^3+T_2K_2^2)+N_tN_r.\mathrm{min}(N_t,N_r))$, where $T$ is the number of AO iterations. 
\section{Communication reliability optimization} \label{Sec:ComReliOpt}
This session focuses on the joint design of RIS phase shifts and beamforming matrix to enhance communication reliability. The objective is to minimize the SER of the detected signal, on which the union bound can be expressed as \cite{JYe2020-RISBER} 
\begin{equation}
    P_S(\mathbf{\Phi}_1,\mathbf{\Phi}_2,\mathbf{F},\mathbf{W}) = \frac{1}{M}\sum_{i=1}^{M}\sum_{j=1,j\neq i}^{M}\mathrm{Pr}\{\mathbf{s}_i\rightarrow \mathbf{s}_j\},
\end{equation}
where $\mathrm{Pr}\{\mathbf{s}_i\rightarrow \mathbf{s}_j\}$ represents the pairwise SER of the symbol $\mathbf{s}_i$ being incorrectly detected as $\mathbf{s}_j$. This probability can be calculated using the squared Euclidean distance as 
$\mathrm{Pr}\{\mathbf{s}_i\rightarrow \mathbf{s}_j\}=Q\left(\sqrt{\frac{d_{i,j}^2(\mathbf{\Phi}_1,\mathbf{\Phi}_2,\mathbf{F},\mathbf{W})}{2\sigma^2}}\right),$ where   $d_{i,j}^2(\mathbf{\Phi}_1,\mathbf{\Phi}_2,\mathbf{F},\mathbf{W}) = ||\mathbf{W}\mathbf{O}\mathbf{F}(\mathbf{s}_i-\mathbf{s}_j)||^2$, and $Q(.)$ denotes the tail distribution function of the standard normal distribution.
The SER minimization problem can be defined as
\begin{align}
     \mathrm{(P4)}: &\underset{\mathbf{F},\mathbf{W},\mathbf{\Phi}_1,\mathbf{\Phi}_2}{\mathrm{minimize}}&& P_S(\mathbf{\Phi}_1,\mathbf{\Phi}_2,\mathbf{F},\mathbf{W}) \\ 
     &\mbox{subject to} &&||\mathbf{F}||_F^2=N_s,\\
     & &&|\Phi_i(j)| = 1, \quad \forall i = 1,2, j = 1,\cdots,K_i.
\end{align}
Obviously, $\mathrm{(P4)}$ is non-convex due to the strong coupling between the variables. Moreover, solving $\mathrm{(P4)}$ will be highly complex, especially when the order of the modulation scheme is high. In this section, we propose the double RISs-aided autoencoder approach that jointly optimizes all the entities in the system as illustrated in Fig.~\ref{fig:End-to-end}, where the transceiver and the double RISs are replaced by neural networks. The SER optimization problem is then viewed as a classical classification problem, in which each symbol in the constellation represents a class. After that, the autoencoder is trained to correctly map received symbols to their corresponding class. It is worth noting that unlike $\mathrm{(P1)}$, in the SER minimization problem, the true labels are always available. Thus, it is reasonable to design a machine learning-based solution for $\mathrm{(P4)}$.
\subsection{Preliminary}
In order to jointly control RISs' phase shifters and precoding matrices, we choose a $1$-dimensional convolutional neural network (1D-CNN) for the proposed framework due to several reasons. First, CNNs can provide a huge gap in computational complexity compared to FCNN \cite{Zhu2019-CNNAE}. Particularly, in a complex system such as double RISs-assisted MIMO, where the number of parameters to be optimized grows exponentially, CNN can greatly reduce the computational complexity of the framework. Second, 1D-CNN with a block length of $L$ can simultaneously encode $L$ data symbols via multiple random channel realizations. With the help of CNN-based encoder, the information is spread over multiple time steps. Thus, the encoded data becomes more robust to the channel fading effects, which results in an improvement in the data decoding performance. Furthermore, unlike FCNN-Autoencoder, the Autoencoder design with a 1D-CNN neural network can adapt to any input length after training, which provides design flexibility.
\subsection{Transmitter Design}
\begin{figure*}
    \centering
    \hspace*{-0cm}
    \includegraphics[trim=0cm 0cm 0cm 0cm, clip=true, width=6.2in]{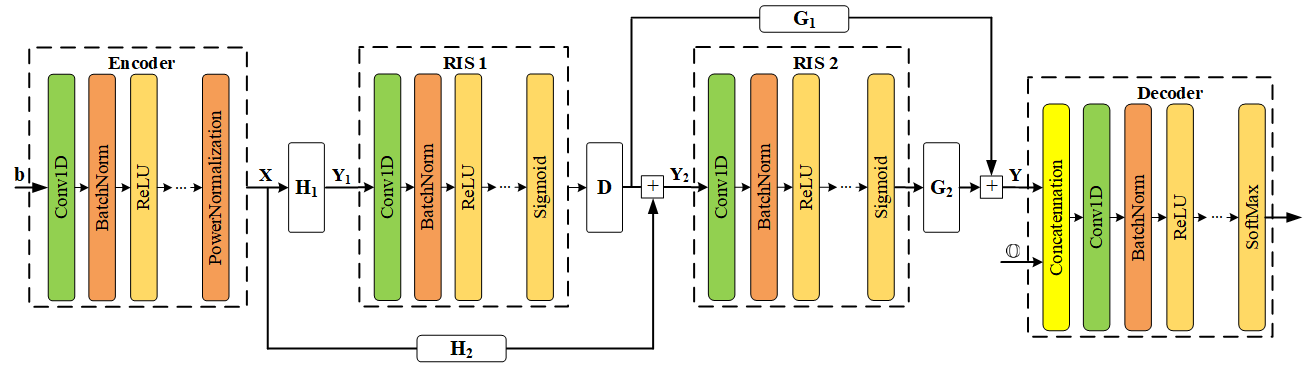}
    \caption{The proposed end-to-end framework where each entity is modeled by a 1D-CNN network.}
    \label{fig:End-to-end}
    \vspace{-0.3cm}
\end{figure*}

In the autoencoder approach, each entity of the system can be replaced with a 1D Convolutional neural network (1D-CNN) as illustrated in Fig.~\ref{fig:End-to-end}. At the transmitter, the data bit stream $\mathbf{b}$ is represented by the one-hot vector with a length of $M$, each corresponding to one of the $M$ possible modulated data signals. The one-hot data are then stacked to a data sequence as $\mathbf{B} = [\mathbf{b}_1,\cdots,\mathbf{b}_L] \in \mathbb{C}^{M \times L}$, with $L$ is the block length of 1D-CNN, to feed to the encoder. The input is then processed by several 1-dimension convolution (Conv1D) layers followed by batch normalization (BN) and activation layers. Before the transmission, the encoded signal is normalized by the power normalization layer, which is a custom layer containing non-trainable parameters. The normalized signal can be calculated as $\mathbf{X} = \frac{P\mathbf{X}'}{\sqrt{\mathbb{E}[|\mathbf{X}'|^2]}}$, where $\mathbf{X} \in \mathbb{C}^{N_t \times L}$ is the output of the encoder, and $\mathbf{X}' \in \mathbb{C}^{N_t \times L}$ is the output of the last 1D-CNN layer. It is worth noting that on the transmitter side, channel information is not required which relaxes the burden of channel estimation and feedback overhead. This is achieved by the fact that 1D-CNN can encode the signal based on the correlation between consecutive symbols. Thus, the encoder can learn to encode data into vectors that are robust to the channel effects.
\subsection{RIS Networks Design}
We use two 1D-CNN models to control the behavior of the double RISs network. In our previous work \cite{HaAn2022-Autoencoder}, the channel information between each link is exploited to optimize the RIS network's parameters. However, in the double RISs-based system, it is challenging to estimate every channel link separately. Therefore, the approach proposed in \cite{HaAn2022-Autoencoder} can not be directly applied to this system. Instead, the received signal at each RIS is fed directly as the input of each RIS model. Specifically, the input of RIS 1 is defined as $\mathbf{Y}_1 = [\mathbf{y}_{11},\cdots,\mathbf{y}_{1L}] \in \mathbb{C}^{K_1 \times L}$, and each element of $\mathbf{Y}_1$ is calculated as $\mathbf{y}_{1i} = \mathbf{H}_1^i\mathbf{x}_i$,
with $i = 1,\cdots L$, the superscript $i$ denotes the channel at the $i$-th symbol. The real part and image part of the input signal is then separated and reshaped into a tensor with the shape of $2K_1 \times L$ and then fed through several 1D-CNN layers attached to BN and ReLU activation layers. The predicted phase shift vector of the RIS 1 is given as $\tilde{\pmb{\Theta}}_1 = [\tilde{\pmb{\theta}}_{11},\cdots,\tilde{\pmb{\theta}}_{1L}] \in \mathbb{C}^{K_1 \times L}$, where $\tilde{\pmb{\theta}}_{1i} = \{\tilde{\theta}_{11}^i,\cdots,\tilde{\theta}_{1K_1}^i\}$ which is followed by the predicted reflection matrix $\tilde{\mathbf{\Phi}}_1^i = \mathrm{diag}(e^{j\tilde{\theta}_{11}^i},\cdots,e^{j\tilde{\theta}_{1K_1}^i})$. 

Similarly, the input of RIS 2 is defined as $\mathbf{Y}_2 = [\mathbf{y}_{21},\cdots,\mathbf{y}_{2L}] \in \mathbb{C}^{K_2 \times L}$, and $\mathbf{y}_{2,i} = (\mathbf{H}_2^i+\mathbf{D}^i\tilde{\mathbf{\Phi}}_1^i\mathbf{H}_1^i)\mathbf{x}_i$, $i = 1,\cdots,L$. With the same structure as RIS model 1, after processing the input, the predicted phase shift vector for RIS is given as $\tilde{\pmb{\Theta}}_2 = [\tilde{\pmb{\theta}}_{21},\cdots,\tilde{\pmb{\theta}}_{2L}] \in \mathbb{C}^{K_2 \times L}$, where $\tilde{\pmb{\theta}}_{2i} = \{\tilde{\theta}_{21}^i,\cdots,\tilde{\theta}_{2K_2}^i\}$. The reflection matrix for RIS 2 at the $i$-th symbol is $\tilde{\mathbf{\Phi}}_2^i = \mathrm{diag}(e^{j\tilde{\theta}_{21}^i},\cdots,e^{j\tilde{\theta}_{2K_2}^i})$. 

\subsection{Decoder Design}
At the receiver side, the received signal and the cascaded channel are exploited for data detection. Firstly, the received signal is calculated based on the current channels and the predicted phase-shift vectors as
\begin{equation}
\begin{split}
     \mathbf{y}_i &= (\mathbf{G}^i_2\tilde{\mathbf{\Phi}}_2^i\mathbf{D}^i\tilde{\mathbf{\Phi}}_1^i\mathbf{H}^i_1 + \mathbf{G}^i_1\tilde{\mathbf{\Phi}}_1^i\mathbf{H}^i_1 + \mathbf{G}^i_2\tilde{\mathbf{\Phi}}_2^i\mathbf{H}^i_2)\mathbf{x}_i\\ 
     &= \mathbf{O}^i\mathbf{x}_i, \quad i = 1,\cdots,L.
\end{split}
\end{equation}

After gathering $L$ received symbols, the received signal $\mathbf{Y} = [\mathbf{y}_1,\cdots,\mathbf{y}_L] \in \mathbb{C}^{N_r \times L}$ and the cascaded channel $\mathbb{O} = \{\mathbf{O}^1;\cdots;\mathbf{O}^L\} \in \mathbb{C}^{N_r\times N_t \times L}$ are concatenated to form a tensor with the shape of $(N_r + N_rN_t) \times L$ which will be used as the input data of the decoder. Finally, the tensor with the size of $2(N_r + N_rN_t) \times L$ comprising the real part and imaginary part of the input will be fed through several 1D-CNN layers followed by BN and ReLU activation layers similar to the encoder. At the end of the decoder, a softmax activation layer is added to cast the output into the tensor $\mathbf{P} = [\mathbf{p}_1,\cdots,\mathbf{p}_L] \in \mathbb{C} ^{M\times L}$, where $\mathbf{p}_i$ is a vector containing the probabilities of all possible messages corresponding to the $i$-th symbol. The decoded message $\hat{\mathbf{B}} = [\hat{\mathbf{b}}_1,\cdots,\hat{\mathbf{b}}_L]$ is determined based on the index of $\mathbf{p}_i$ with the highest probability. Overall, the parameter settings for the end-to-end system are given in Table~\ref{CNN-Structure}.
\subsection{Optimization Process}
The parameters of the encoder, decoder, and RIS models are jointly optimized by minimizing the loss function with a given modulation scheme as
\vspace*{-0.1cm}
\begin{equation} \label{Pro:Loss}
\begin{aligned}
& \underset{\{f,g,r_1,r_2\}·}{\mathrm{minimize}} && \mathcal{L}_{\mathrm{AE}} (\psi_f,\psi_{g},\psi_{r_1},\psi_{r_2}),\\
& \mbox{subject to} && \mathbf{b} \in \mathcal{M}, 
\end{aligned}
\vspace*{-0.1cm}
\end{equation}
where $\mathcal{M}$ is the finite constellation set defined by the M-QAM, $\psi_f, \psi_{g}, \psi_{r_1}, $ and $\psi_{r_2}$ are the parameters of the encoder $f$, the decoder $g$, and the double RIS network $r_1$ and $r_2$, respectively. We notice that the data detection of the transmitted bits can be regarded as a typical classification problem. Thus, the binary cross-entropy loss function is readily used for the optimization. The loss function $\mathcal{L}_{\mathrm{AE}} (\psi_f,\psi_{g},\psi_{r_1},\psi_{r_2})$ is given as
\begin{equation}\label{eq:Loss}
\begin{split}
    &\mathcal{L}_{\mathrm{AE}} (\psi_f,\psi_{g},\psi_{r_1},\psi_{r_2}) \\
    &= \frac{1}{LM}\sum_{i=0}^{L-1}\sum_{m=0}^{M-1}-\Big\{[\mathbf{p}_i]_m\log[\mathbf{b}_i]_m\\
    &+(1-[\mathbf{p}_i]_m)\log(1-[\mathbf{b}_i]_m)\Big\}.
\end{split}
\end{equation}

The update of parameters can be done by applying a gradient descent algorithm and back propagation procedure on \eqref{eq:Loss}. In order to demonstrate the ability to optimize the end-to-end system, we define the full network function as $F(\mathbf{B},\zeta)$, where $\zeta$ is the parameter of the end-to-end system to be optimized. For two tensors $\mathbb{M} = [\mathbf{M}^1,\cdots,\mathbf{M}^L]$, and $\mathbb{N} = [\mathbf{N}^1,\cdots,\mathbf{N}^L]$,  the operator $\bullet$ is denoted as $\mathbb{M} \bullet \mathbb{N} = [\mathbf{M}^1\mathbf{N}^1, \cdots, \mathbf{M}^L\mathbf{N}^L]$. Based on the forward procedure described in the aforementioned subsection, the network function can be written as 
\begin{equation} \label{eq:systemFunction}
\begin{split}
    F(\mathbf{B},\zeta) &= g\Big(\mathbb{G}_2\bullet r_2\big(\mathbb{D} \bullet  r_1(\mathbb{H}_1\bullet f(\mathbf{B},\psi_f),\psi_{r_1})\\
    +&\mathbb{H}_2\bullet f(\mathbf{B},\psi_f),\psi_{r_2}\big)+\mathbb{G}_1\bullet r_1\big(f(\mathbf{B},\psi_f),\psi_{r_1}\big),\psi_g\Big),    
\end{split}
\end{equation}
where $\mathbb{H}_i = [ \mathbf{H}_i^1,\cdots ,\mathbf{H}_i^L ]$, $\mathbb{G}_i = [ \mathbf{G}_i^1,\cdots ,\mathbf{G}_i^L ]$, and $\mathbb{D} = [ \mathbf{D}^1,\cdots ,\mathbf{D}^L ]$
\footnote{Even though channel information of each channel link is required in the training phase, in the inference phase, the RISs and receiver networks, i.e $r_1, r_2$, and $g$, only take received signals as the input. Therefore, the proposed end-to-end framework can release the burden of channel estimation in the online testing phase. }. 
From \eqref{eq:systemFunction}, we can observe that the forward and backward gradient can be readily computed on $\mathbf{F(\mathbf{B},\zeta)}$ without encountering recursive problems.
\begin{table}
	    \centering
    	    \caption{Parameter setting for the encoder, decoder, and RIS models.}
    	\begin{tabular}{ | P{2cm} | P{3cm}| P{2.5cm}|} 
          \hline
          Layers & Parameters  & Output dimensions \\ 
          \hline
          \multicolumn{3}{|c|}{Encoder}\\ 
          \hline
          Input & None & $M\times L$\\
          \hline
          Conv1D + BN + ReLU & kernel = 1, filter = 256 & $256 \times L$\\
          \hline
          Conv1D + BN + ReLU & kernel = 1, filter = 256 & $256 \times L$\\
          \hline
          Conv1D + BN  & kernel = 1, filter = $2N_t$ & $2N_t \times L$\\
          \hline
    
          \multicolumn{3}{|c|}{Decoder} \\ 
          \hline
          Input & None & $(2N_r+2N_tN_R)\times L$ \\
          \hline
          Conv1D+ BN + ReLU & kernel = 1, filter = 512 & $512 \times L$\\
          \hline
          Conv1D + BN + ReLU & kernel = 1, filter = 512 & $512 \times L$\\
          \hline
          Conv1D + solfmax & kernel = 1, filter = $M$ & $M \times L$\\
          \hline
          \multicolumn{3}{|c|}{RIS model $i, i = 1,2$}\\
          \hline
          Input & None & $2K_i\times L$ \\
          \hline
          Conv1D + BN + ReLU & kernel = 1, filter = 512 & $512 \times L$\\
          \hline
          Conv1D + BN + ReLU & kernel = 1, filter = 512 & $512 \times L$\\
          \hline
          Conv1D  & kernel = 1, filter = $K_i$ & $K_i \times L$\\
          \hline

        \end{tabular} \label{CNN-Structure}
    \end{table}
\subsection{Computational complexity}
As reported in \cite{Serkan2021-1DCNN}, the complexity of the 1D-CNN model can be calculated as $\mathcal{O}(\sum_{n=1}^{N}Lk_n^2F_{n-1}F_n)$, where $N$ is the number of layer, $L$ is the block length, $k_n$ is the kernel size of the $n$-th layer, and $F_{i}$ is the number of filters in the $i$-th layer. In general, the complexity of 1D-CNN increases with the block length as well as the kernel size. In this paper, to demonstrate the complexity in terms of the MIMO system parameters, we assume that the kernel size, the number of layers, and the number of filters are constants. Therefore, the computational complexities for the encoder, decoder, and RIS models scale like of $\mathcal{O}(LMN_t)$, $\mathcal{O}(L(N_r+N_rN_t)M)$, and $\mathcal{O}(L(K_1^2+K_2^2))$, respectively. Overall, the complexity of the end-to-end system can be evaluated as $\mathcal{O}(L(MN_rN_t+K_1^2+K_2^2))$.
\section{Simulation results} \label{Sec:SiRe}
In this section, we provide simulation results to evaluate the performance of the considered double-RIS-aided point-to-point system along with the proposed approaches.
\subsection{Simulation Settings}
\begin{figure}
    \centering
    \includegraphics[trim=0cm 0cm 0cm 0cm, clip=true, width=2.5in]{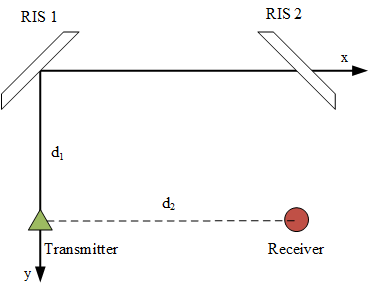}
    \caption{System setup for the double-RIS system (top view).}
    \label{fig:SystemSetup}
    \vspace{-0.5cm}
\end{figure}

We apply a three-dimensional (3D) Cartesian coordinate system to illustrate the location of the entities in the considered system. We assume that both the transmitter and RIS surfaces are located at the same altitude above the receiver by $d_H$ meter (m). The location of the central point of the transmitter, receiver, RIS 1 and RIS 2 are set as ($0$, $d_1$, $d_H$), ($d_2$, $d_1$, $0$), ($0$, $0$, $d_H$), and ($d_2$, $0$, $0$), respectively. In Fig.~\ref{fig:SystemSetup}, we illustrate the horizontal projections of the system setup. As shown in Fig.~\ref{fig:SystemSetup}, we set the azimuth angles of RIS 1 and RIS 2 with respect to the x-axis as $\pi/4$ and $3\pi/4$, respectively. The details of AoA/AoD angles for all channel links as well as their LoS paths are summarized in Table.~\ref{Channel-Para}. The distance path loss model is given following the 3GPP propagation environment \cite{Huayan2020-WSR-RIS} as $\gamma = 35.6 + 22\log (d)$, where d is the individual link distance in meter. Unless specified otherwise, the other system parameters are set as $P = 30 $dBm, $\sigma^2 = -90 $dBm and $\epsilon = 10^{-5}$ for the simulation. And finally, a MIMO system model with $N_t = N_r = 16$ is considered for the simulation.

\begin{table*}
    \centering
    \caption{Summary of channel parameters used in the simulation}
    \begin{tabular}{ | P{1.5cm} | P{2cm}|P{2.6cm}|P{2.7cm}|P{2.5cm}|P{3.6cm}|} 
      \hline
      Channel link& $\mathbf{H}_1$  &  $\mathbf{H}_2$  &$\mathbf{G}_1$&$\mathbf{G}_1$&$\mathbf{D}$\\ [0.15cm]
      \hline
      Distance (m) & $d_1$ & $\sqrt{d_1^2+d_2^2}$ & $\sqrt{d_1^2+d_2^2+d_H^2}$ & $\sqrt{d_1^2+d_H^2}$ &$d_2$\\[0.2cm] 
      \hline
      AoA &$\theta^\mathrm{A}_\mathrm{T1} = \frac{\pi}{4}$, $\phi^A_{T1}=0$&$\theta^\mathrm{A}_\mathrm{T2}=\frac{\pi}{4}-\mathrm{arctan}(\frac{d_2}{d_1})$, $\phi^A_{T2}=0$&$\theta^A_\mathrm{R1}=\mathrm{arctan}(\frac{d_2}{d_1})$&$\theta^A_\mathrm{R2}=0 $&$\theta^A_\mathrm{R}=\frac{\pi}{4}$, $\phi^A_\mathrm{R}=0$\\[0.5cm] 
      \hline
      AoD &$\theta^D_\mathrm{T1}=\frac{\pi}{2}$&$\theta^D_\mathrm{T2}=\mathrm{arctan}(\frac{d_1}{d_2})$&$\theta^D_\mathrm{R1}=\mathrm{arctan}(\frac{d_1}{d_2})$, $\phi^D_\mathrm{R1}=\mathrm{arctan}(\frac{d_H}{\sqrt{d_1^2+d_2^2}})$&$\theta^D_\mathrm{R2}=\frac{\pi}{4}$, $\phi^D_\mathrm{R2}=\mathrm{arctan}(\frac{d_H}{d_1})$&$\theta^D_\mathrm{R}=\frac{\pi}{4}$, $\phi^D_\mathrm{R}=0$\\[0.5cm] 
      \hline
     LoS link &\multicolumn{2}{c|}{$\mathbf{\bar{H}}_i=\mathbf{a}_i(\theta^A_\mathrm{Ti},\phi^A_\mathrm{Ti})\mathbf{a}_T(\theta^D_\mathrm{Ti})$}&\multicolumn{2}{c|}{$\mathbf{\bar{G}}_i=\mathbf{a}_R(\theta^A_\mathrm{Ri})\mathbf{a}_i(\theta^D_\mathrm{Ri},\phi^D_\mathrm{Ri})$}&$\mathbf{\bar{D}}=\mathbf{a}_2(\theta^A_\mathrm{R},\phi^A_\mathrm{R})\mathbf{a}_1(\theta^D_\mathrm{R},\phi^D_\mathrm{R})$\\[0.15cm]
      \hline
    \end{tabular} \label{Channel-Para}
\end{table*}

In the training phase, we generate $100,000$ different data symbols as well as channel realizations as the dataset. Among them, we use $90,000$ data symbols for the training phase and $10,000$ symbols for the test phase. In the training phase, all the 1D-CNN networks are jointly optimized using ADAM optimizer\cite{Adam-2014}, where the initial learning rate is set as $0.001$, and is decayed by a factor of 5 after every 5 epochs. With the help of BN layers, convergence is achieved quickly, thus, 20 epochs were used for the training phase. 
%We summarize the training setup in Table~\ref{Training_Para}. 
During the training phase, the power of noise, i.e, the SNR of the transmitted signal, is varied in different system setups. 

Furthermore, all the proposed schemes are implemented on a computer with an Intel Core i5-10400 CPU @2.90 GHz, an NVIDIA Geforce GTX 1050 TI 16GB memory. Matlab 2021a is used for the Monte-Carlo simulations while Python 3.8.3 and Pytorch library is used to implement neural network model. In the following subsections, we evaluate the performance of the following designs: $i)$ \textit{double RIS-assisted Joint Transmitter and Receiver Design} as presented in Section \ref{Sec:ChanCapOpt}, and it is denoted as ``Model-based" in the figures;
$ii)$ \textit{double RIS-assisted Autoencoder} as presented in Section \ref{Sec:ComReliOpt}, and it is denoted as ``Autoencoder" in the figures.

\subsection{Power Scaling Law}
\begin{figure}
    \centering
    \includegraphics[trim=0cm 0cm 0cm 0cm, clip=true, width=4.5in]{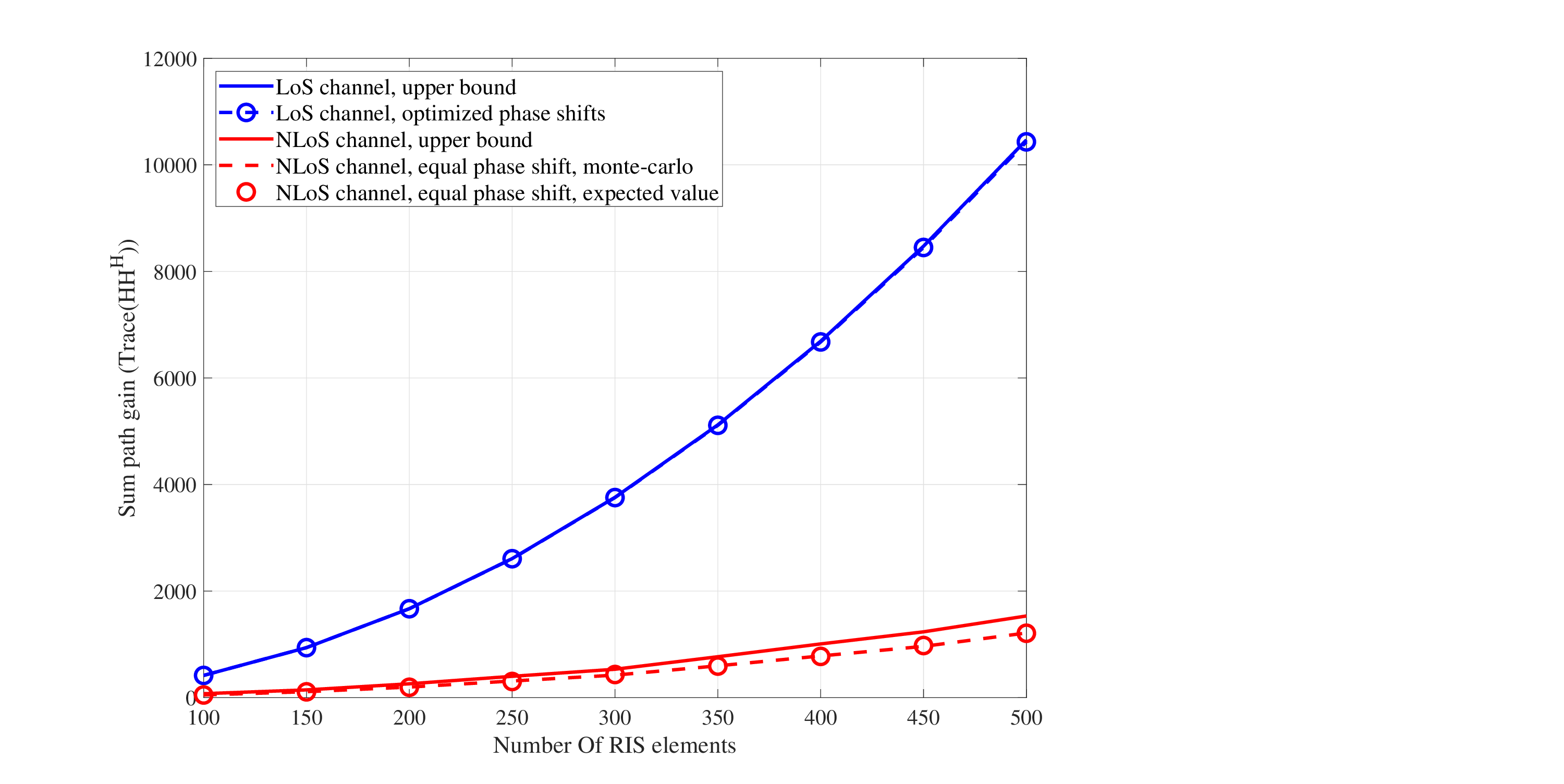}\hspace*{-3cm} 
    \caption{The upper bound for system channel power gain versus number of RIS  reflecting elements in LoS and NLoS channel cases, with $\mathsf{SC} = 3$.}
    \label{fig:BoundNRIS}
\vspace*{-0.8cm}
\end{figure}

In this subsection, we study the impact of setup parameters on the system performance based on the results presented in Lemma \ref{Lemma1}. The location of the transceiver, as well as RISs, are set as $d_1 = 100$ m, $d_2 = 200$ m, $d_H = 2$ m. Firstly, we plot the expected value of $\mathrm{tr}(\mathbf{O}\mathbf{O}^H)$ obtained from \eqref{eq:ChannelSecondMoment} and its average value with the monte-carlo method over 1,000 channel realizations in Fig.~\ref{fig:BoundNRIS}. As can be observed, the average value converges to the expected one calculated in \eqref{eq:ChannelSecondMoment} that verifies the result in Lemma \ref{Lemma1}. Moreover, we illustrate the upper bounds of the cascaded channel power gain for NLoS and LoS channels as calculated in \eqref{eq:PowerLaw} and \eqref{eq:NLoSPowerLaw}. To show the tightness of these bounds, the upper bound in NLoS channel case is compared with the expected value of $\mathrm{tr}(\mathbf{O}\mathbf{O}^H)$ with equal phase shifts. For LoS channel case, since channels are constants, we compare the upper bound with the value of $\mathrm{tr}(\mathbf{O}\mathbf{O}^H)$ with optimized phase shifts. We can observe from Fig.~\ref{fig:BoundNRIS} that the upper bounds are very tight as the number of reflecting elements increases in both cases. Furthermore, the upper bound for LoS channel increases much faster than that of NLoS channel since the bound for LoS channel scales with the number of reflecting elements in order of $\mathcal{O}(K_1^2K_2^2)$ while for NLoS channel, it scales in the order of $\mathcal{O}(K_1K_2)$.
\vspace{-0.3cm}
\subsection{Capacity Evaluation}
\begin{figure*}
\centering
\begin{subfigure}{0.45\textwidth}
    \centering
    \includegraphics[trim=0cm 0cm 0cm 0cm, clip=true, width=4.5in]{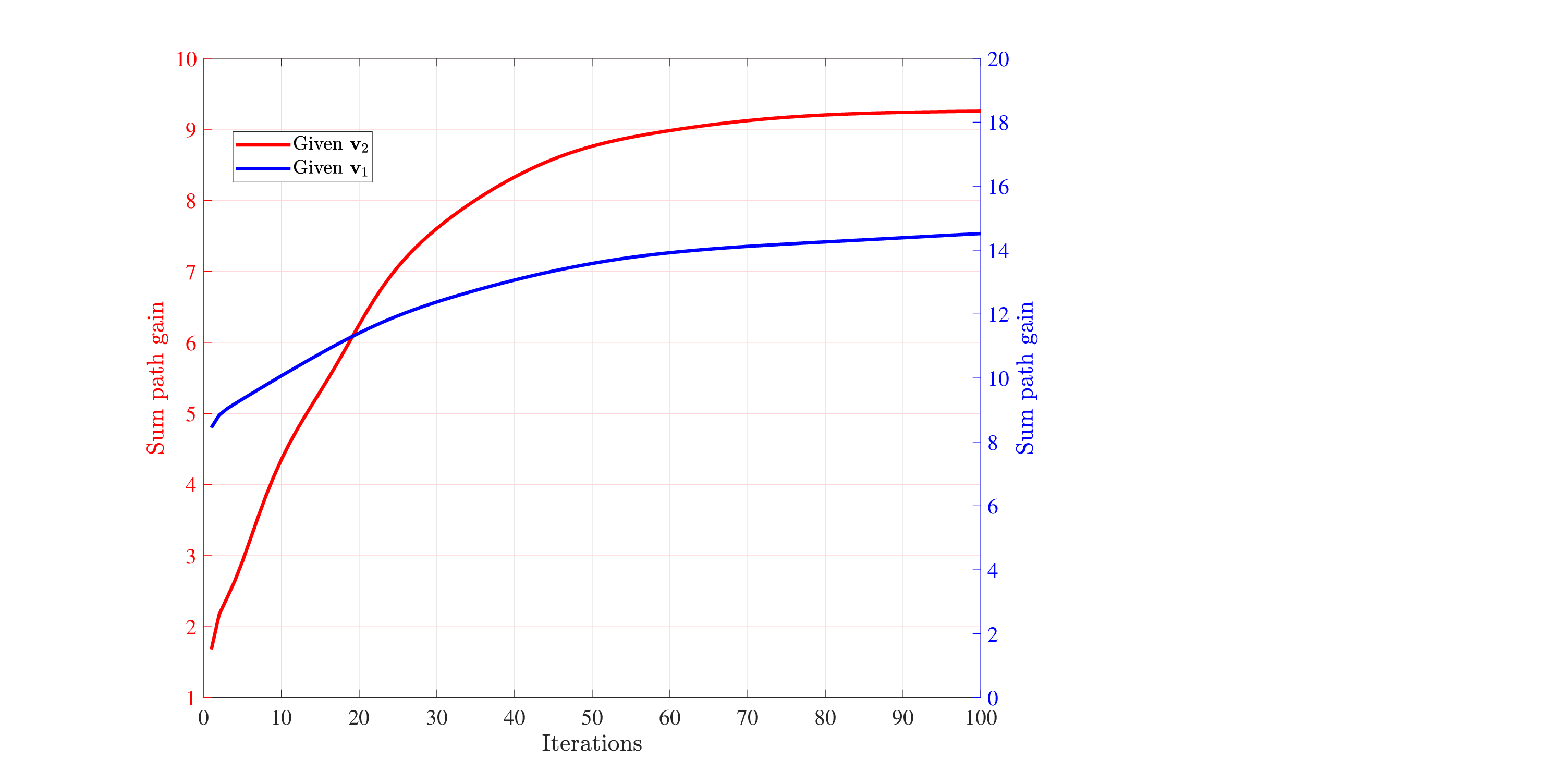}
    \caption{Convergence behavior of ADMM framework for phase shift optimization.}
    \label{fig:InnerConvergence}
\end{subfigure}
\hfill
\begin{subfigure}{0.45\textwidth}
\centering
    \includegraphics[trim=0cm 0cm 0cm 0cm, clip=true, width=4.5in]{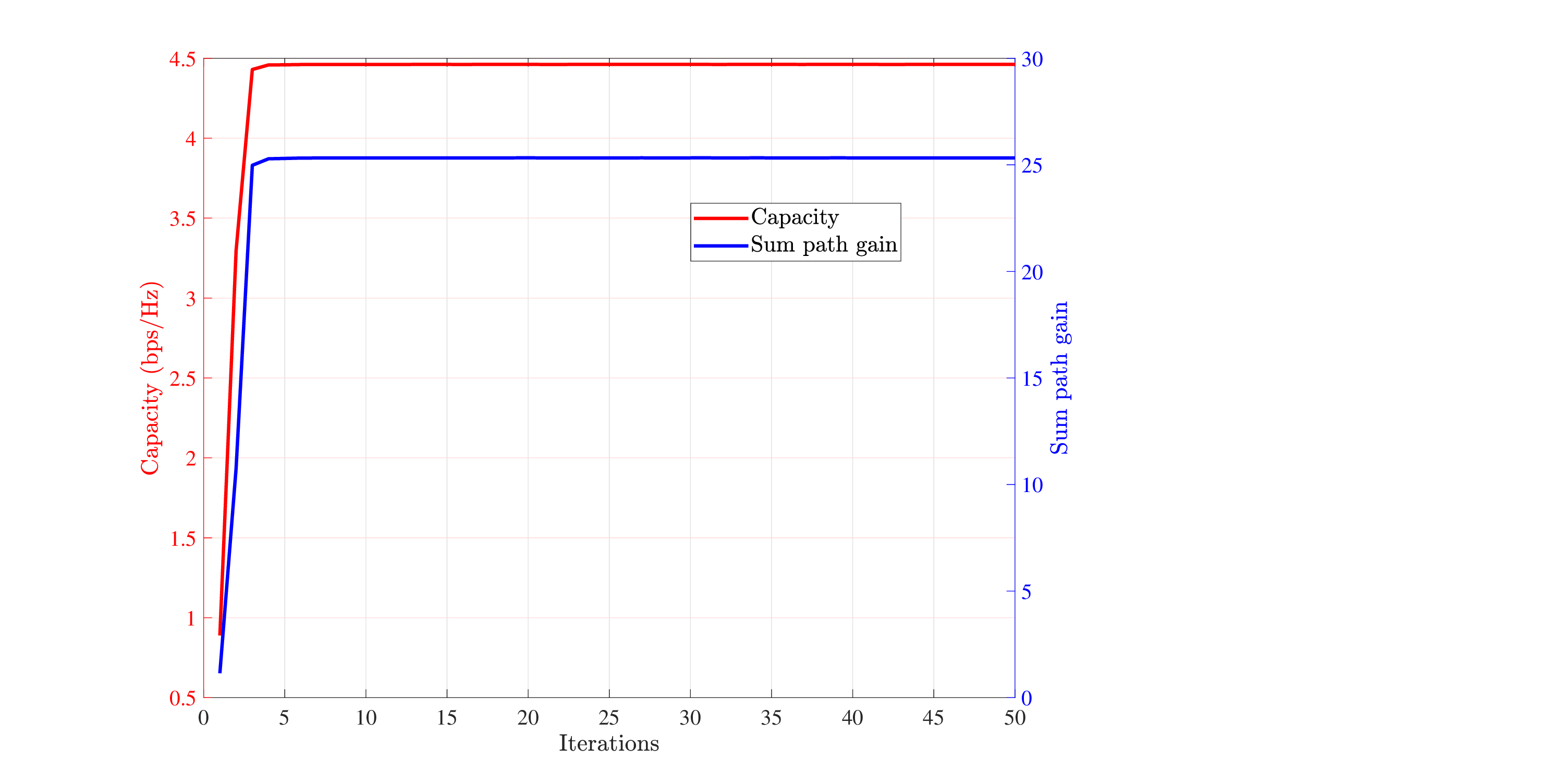}
    \caption{Convergence behavior of the proposed AO framework over 1000 channel realization.}
    \label{fig:AOConvergence}
\end{subfigure}
\caption{Convergence behavior of Algorithm 1.}
\label{fig:Convergence}
\vspace{-0.5cm}
\end{figure*}

In this subsection, we analyze the capacity of the proposed system under the model-based approach. We first evaluate the convergence behavior of Algorithm 1. The channel parameters are set as $\epsilon_i = \delta_i = \mu = 0$, and $\mathsf{SC}_{\mathbf{A}} = 15, \mathbf{A} \in \{\mathbf{H}_i,\mathbf{G}_i,\mathbf{D}\}, i = 1,2$. The convergence behavior of Algorithm 1 is illustrated in Fig.~\ref{fig:Convergence} over 1,000 channel realizations. First, in Fig.~\ref{fig:InnerConvergence}, we show the convergence behavior of ADMM framework for solving (P3.1) and (P3.2) in the first outer iteration. As can be seen, the ADMM optimization frameworks for both (P3.1) and (P3.2) converge after about 100 iterations. Next, we illustrate the optimization convergence of the proposed AO framework in Fig.~\ref{fig:AOConvergence}. In this figure, to show the equivalence between $\mathrm{(P3)}$ and $\mathrm{(P1)}$, we plot the objective of $\mathrm{(P3)}$ in each iteration, and the objective of $\mathrm{(P1)}$ given variables obtained from solving $\mathrm{(P3)}$.  As can be observed, Algorithm 1 converges monotonically and the speed of convergence is fast (the average number of iterations needed to reach the convergence is 5). Moreover, the convergence behavior of the capacity is the same as that of sum path gain of the channel, which validates the equivalence between $\mathrm{P(1)}$ and $\mathrm{(P3)}$.

\begin{figure*}
\centering
\begin{subfigure}{0.45\textwidth}
    \centering
    \includegraphics[trim=0cm 0cm 0cm 0cm, clip=true, width=4.5in]{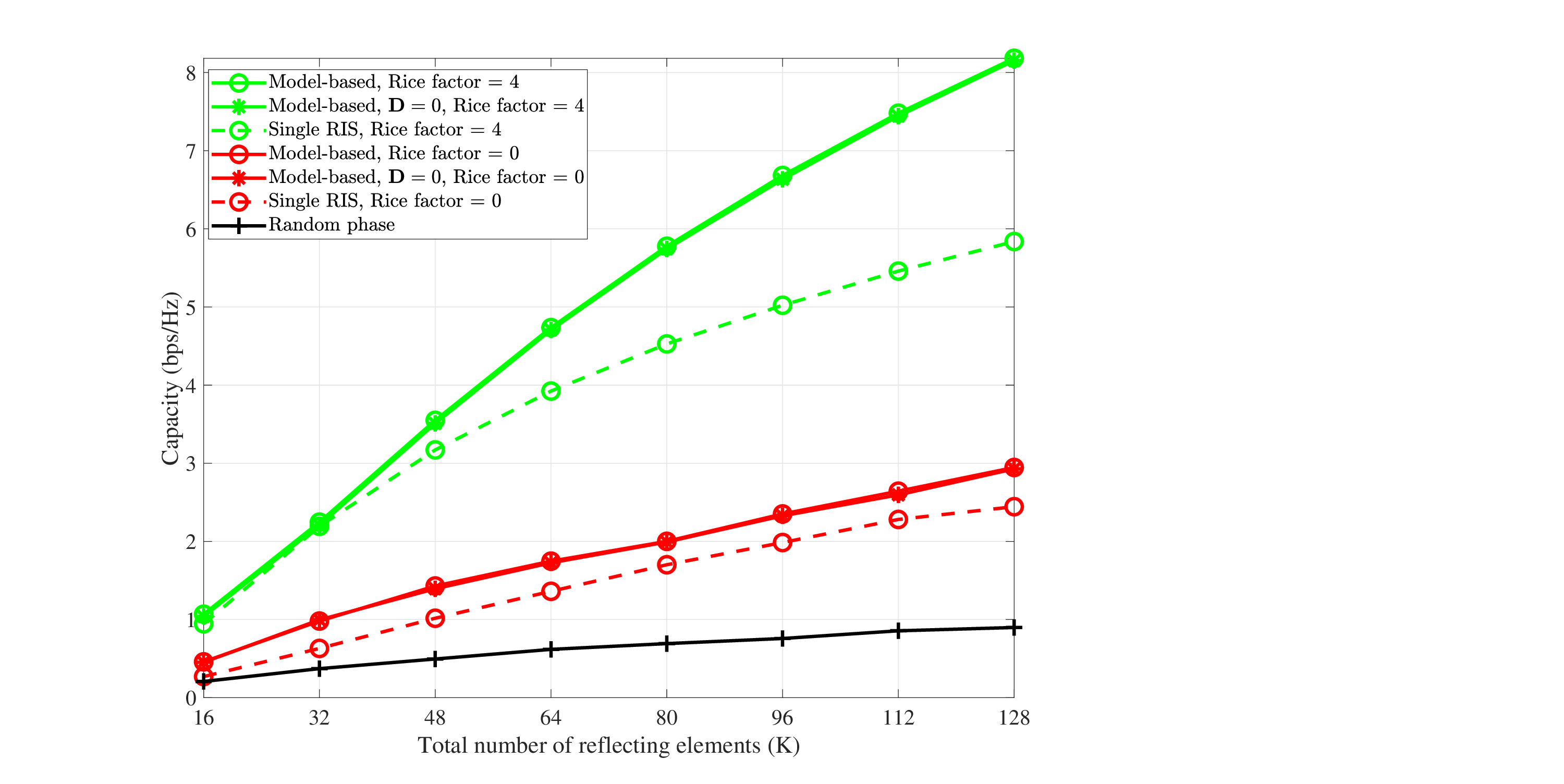}
    \caption{Channel capacity performance with $d_1 = 100$ m, $d_2 = 200$ m.}
    \label{fig:CapacityLongd}
\end{subfigure}
\hfill
\begin{subfigure}{0.45\textwidth}
\centering
    \includegraphics[trim=0cm 0cm 0cm 0cm, clip=true, width=4.5in]{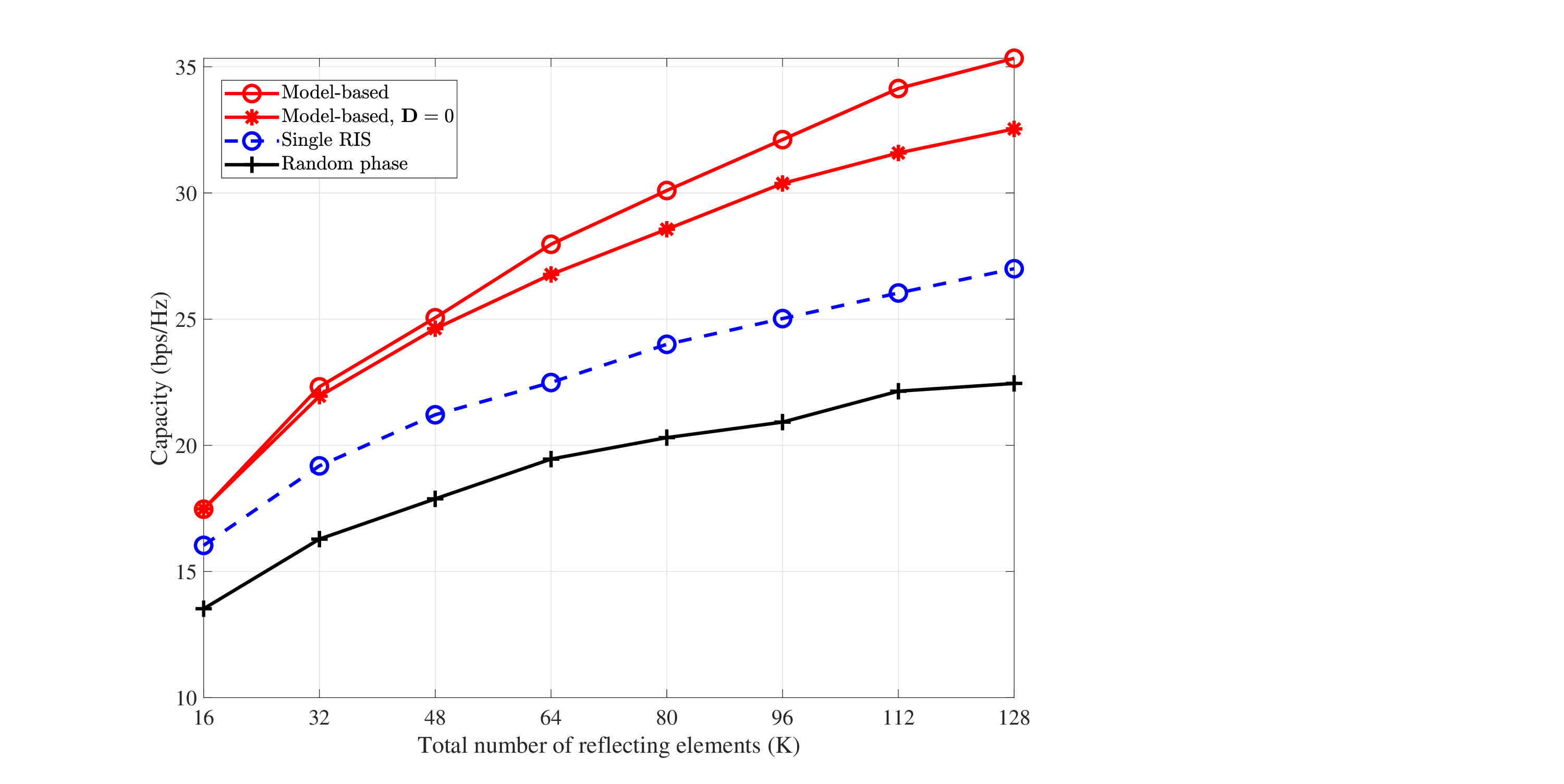}
    \caption{Channel capacity performance with $d_1 = 20$ m, $d_2 = 5$ m.}
    \label{fig:CapacityShortd}
\end{subfigure}
 
\caption{Channel capacity [bps/Hz] versus the number of  reflecting elements at each RIS.}
\label{fig:Capacity}
\vspace{-0.2cm}
\end{figure*}

Next, we demonstrate the superior capacity of our model-based approach according to the number of RIS elements in Fig.~\ref{fig:Capacity}. Additionally, we present the performance of a single-RIS assisted system to highlight the capacity gains achieved with double-RISs. As shown in Fig.~\ref{fig:CapacityLongd}, the capacity of both double- and single-RIS assisted systems increases with larger Rician factors, validating the results of Lemma 1. Furthermore, the double-RIS model outperforms the single-RIS model, emphasizing the advantage of using double-RISs over their single-RIS counterparts. Specifically, in the high Rician factor regime, the gap between the two models increases as the number of reflecting elements grows, which is consistent with the difference in their power scaling orders ($\mathcal{O}(K^4)$ and $\mathcal{O}(K^2)$)\cite{YHan2020-DoubleIRS}, with $K = K_1+K_2$ is the total number of RIS reflecting elements. Furthermore, if we exclude the double-reflection link, i.e. $\mathbf{D} = 0$, the system capacity remains relatively unchanged. This is not surprising given that the double-reflection link becomes insignificant when the distance pathloss is high. However, as shown in Fig.~\ref{fig:CapacityShortd}, in a shorter-distance setup,  absence of  the double-reflection link decreases the performance, particularly when the number of RIS elements becomes large, particularly when the number of RIS elements becomes large.
% In the high-Rician-factor regime, the power law of the system is in order of $\mathcal{O}(K_1^2K_2^2)$ while the power scale in the low-Rician-factor is in order of $\mathcal{O}(K_1K_2)$. 

\vspace{-0.0cm}
\vspace{-0.0cm}
\begin{figure*}[t]
\centering
\begin{subfigure}{0.45\textwidth}
    \centering
    \includegraphics[trim=0cm 0cm 0cm 0cm, clip=true, width=4.5in]{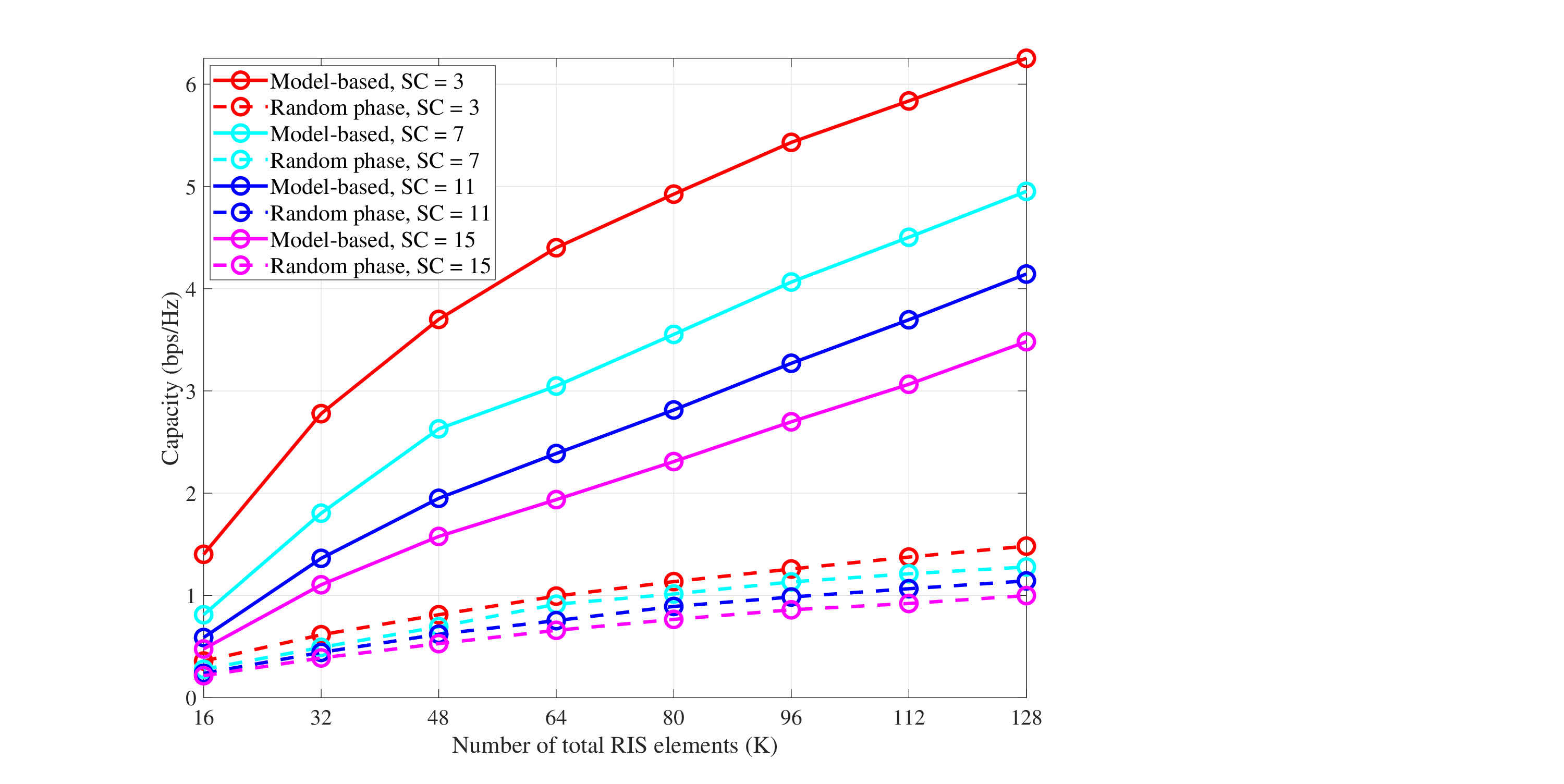}
    \caption{Low-SNR regime, with $d_1 = 100$ m, $d_2 = 200$ m.}
    \label{fig:ImpactScatter}
\end{subfigure}
\hfill
\begin{subfigure}{0.45\textwidth}
\centering
    \includegraphics[trim=0cm 0cm 0cm 0cm, clip=true, width=4.5in]{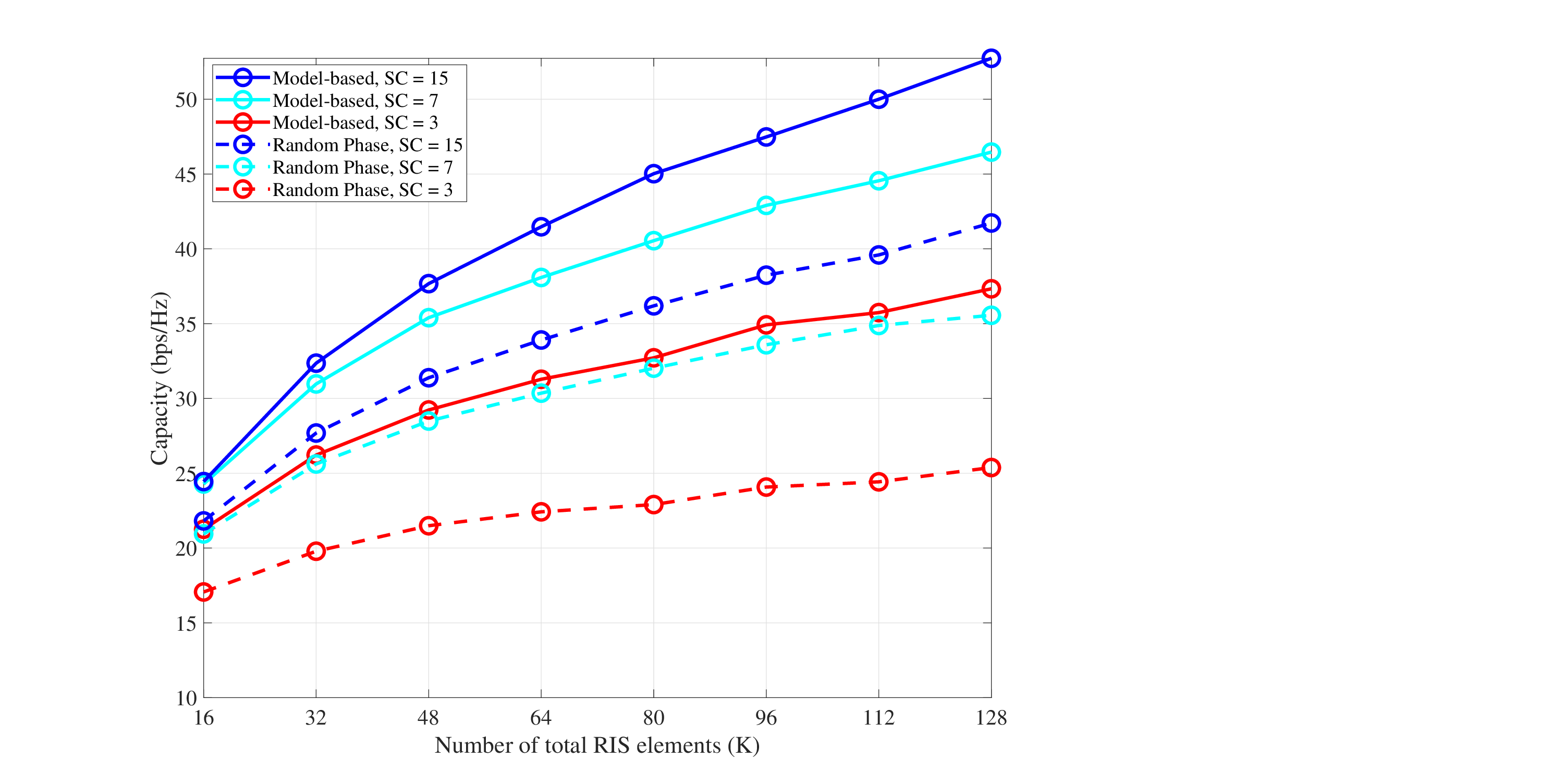}
    \caption{High-SNR regime, with $d_1 = 5$ m, $d_2 = 50$ m.}
    \label{fig:ImpactScatterHighPow}
\end{subfigure}
 
\caption{Impacts of the number of scatterers to the channel capacity.}
\label{fig:figures}
\vspace{-0.5cm}
\end{figure*}
\vspace{-0.0cm}
We further evaluate the impact of the number of scatterers between each link  on the capacity of the model-based approach in Fig.~\ref{fig:figures}. In this setup, we set the Rician factors as $ \epsilon_i = \delta_i = \mu = 0$, and the number of scatterers in each link is equally set as $\mathsf{SC}_{\mathbf{H}i} = \mathsf{SC}_{\mathbf{G}i} = \mathsf{SC}_{\mathbf{D}} = \mathsf{SC}$. Moreover, we consider two sets of locations given by $d_1 = 100$ m, $d_2 = 200$ m and $d_1 = 5$ m, $d_2 = 50$ m, corresponding to the low-SNR regime and high-SNR regime, respectively. The performance of the examined schemes in the low-SNR regime is first illustrated in Fig.~\ref{fig:ImpactScatter}. As is evident, in both the random phase-shift scheme and optimized phase-shift scheme, when the number of scatterers increases, a drop in channel capacity is observed. This result is expected since it has been shown in \eqref{eq:NLoSPowerLaw} that the sum-path-gain of the cascaded channel will decrease with larger number of scatterers. However, we also show in Fig.~\ref{fig:ImpactScatterHighPow} that in high-SNR regime, the MIMO system enjoys the multiplexing gain when channel rank is increased. Therefore, it is shown in Fig.~\ref{fig:ImpactScatterHighPow} that even though the channel energy is decreased with larger number of scatterers, the channel capacity can still be enhanced since multiplexing gain is more important than power gain in high SNR regime.
\vspace{-0.3cm}
\subsection{Symbol Error Rate Evaluation}
In this subsection, we demonstrate the symbol error rate performance of a double RIS-assisted MIMO system under an end-to-end approach. For this evaluation, we set the location of the system as $d_1 = 100$ m, $d_2 = 200$ m, $d_H = 2$ m, and the channel parameters are set as $\epsilon_i = \delta_i = \mu = 4$, and $\mathsf{SC}_{\mathbf{A}} = 15, \mathbf{A} \in \{\mathbf{H}_i,\mathbf{G}_i,\mathbf{D}\}, i = 1,2$. 

\begin{figure*}
    \centering
    %\hspace{-0.2cm}
    \begin{subfigure}{0.45\textwidth}
        \centering
         \includegraphics[trim=0cm 0cm 0cm 0cm, clip=true, width=4.5in]{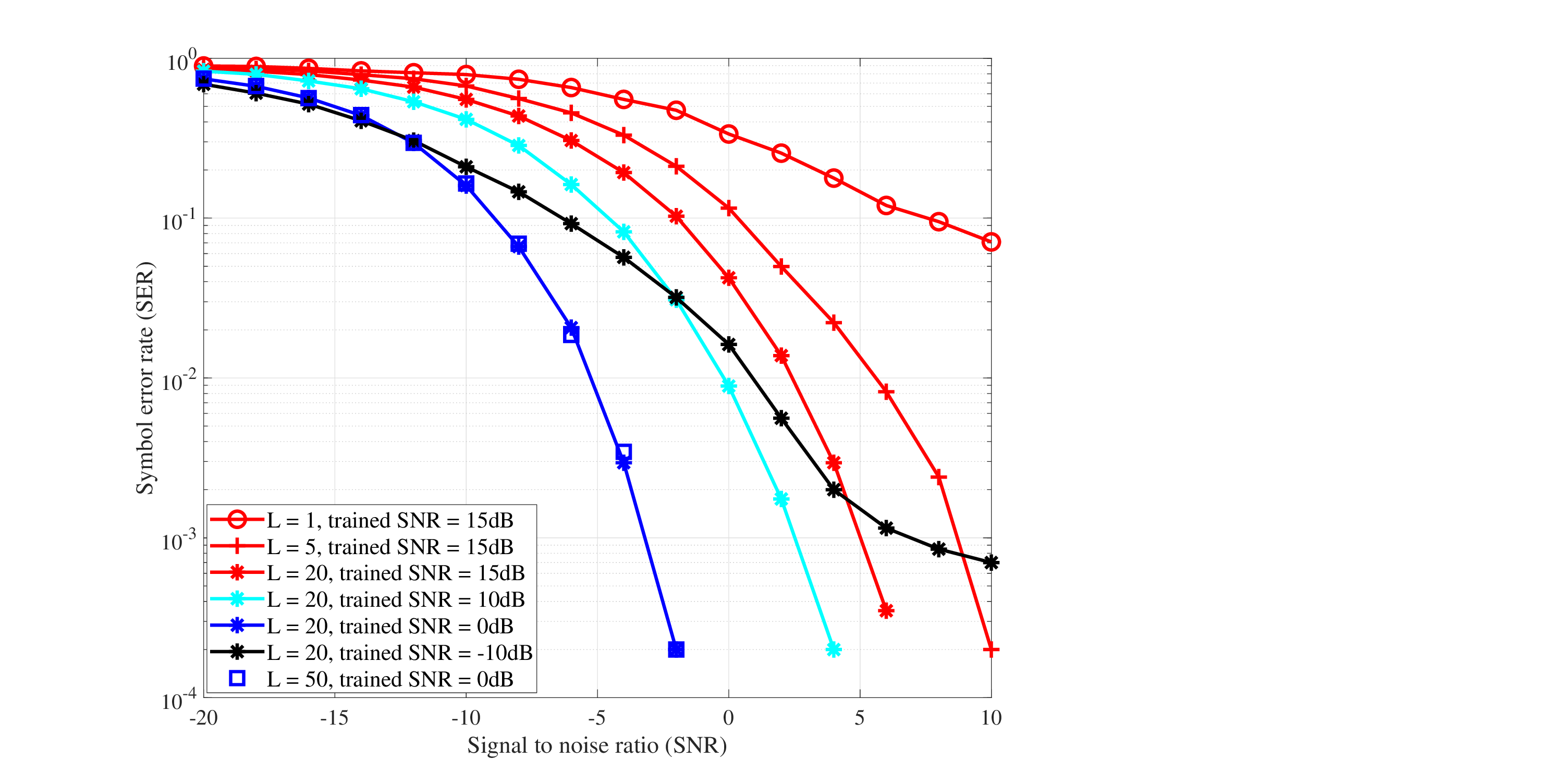}
        \caption{Performance of Autoencoder with 16 QAM.}
        \label{fig:ImpactSNR}   
    \end{subfigure}
    \hfill
    \begin{subfigure}{0.45\textwidth}
        \centering
        %\hspace{-0.2cm}
        \includegraphics[trim=0cm 0cm 0cm 0cm, clip=true, width=4.5in]{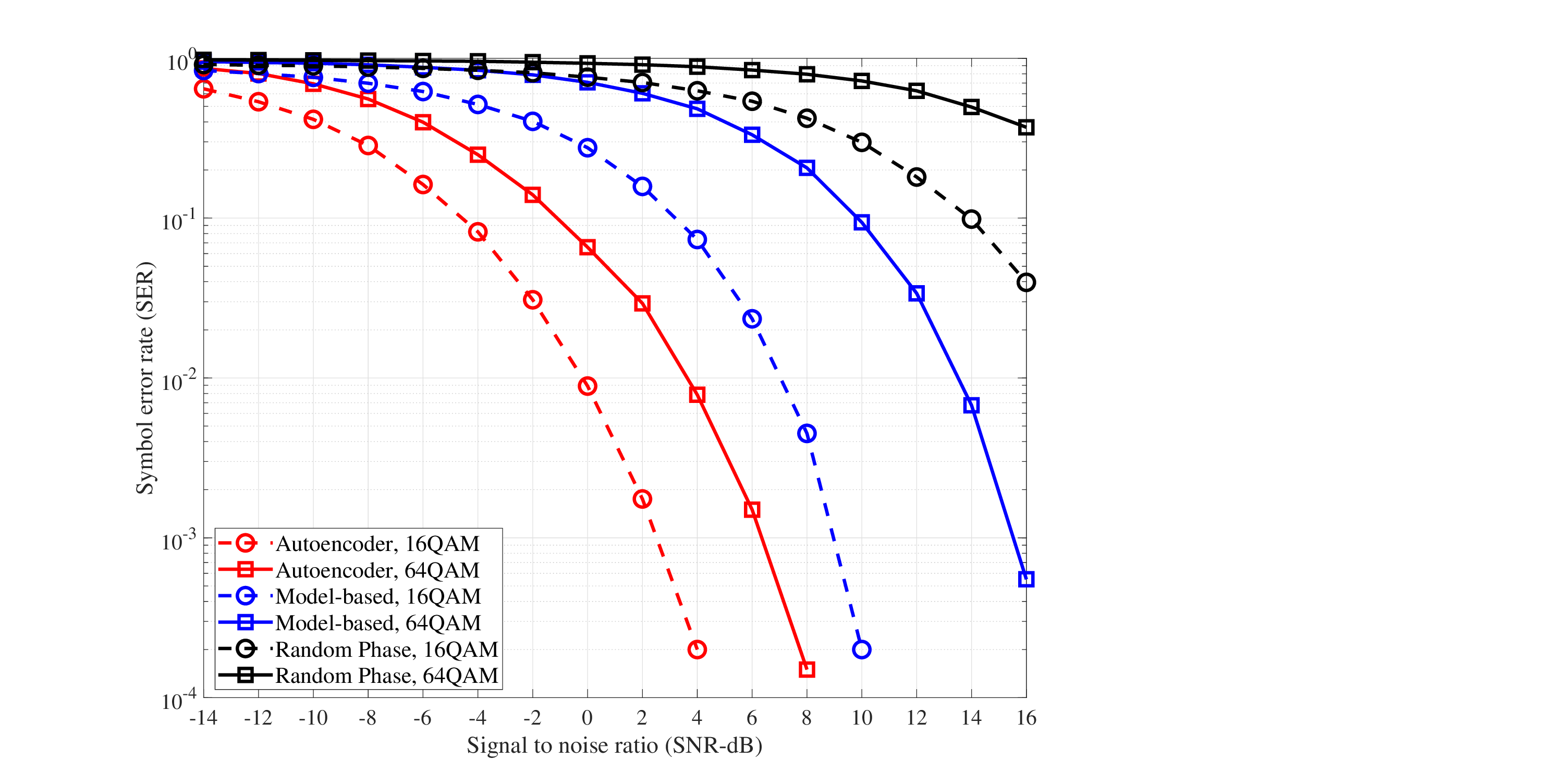}
        \caption{Comparison with other benchmark schemes.}
        \label{fig:SERCompare}
    \end{subfigure}
    \caption{The detection performance of Autoencoder.}
    \vspace{-0.4cm}
\end{figure*}

We first  examine the impact of block length and SNR values by training the end-to-end system with varying values of both parameters. As illustrated in Fig~\ref{fig:ImpactSNR}, the reliability of the system can be enhanced by utilizing a larger block length as shown in \cite{Jiang2019-turboAE}. Specifically, by jointly encode $L$ symbols over random channel realizations during the training phase, transmitting symbols become robust to the  random channel effect, leading to an enhancement in SER performance. This result shows the superiority of 1D-CNN autoencoder compared to a simple FCNN-based design. Furthermore, we also observe that when the noise power increases in the training phase, the SER performance of the "Autoencoder" improves significantly. This can be explained by the fact that increasing the noise power in the training phase will force the "Autoencoder" to optimize the system in a way that is robust to the impact of noise. Consequently, the SER performance is greatly boosted. However, as shown in Fig.~\ref{fig:ImpactSNR}, when the noise power is too high, i.e. SNR = -10dB, Autoencoder fails to detect transmitted symbols since high power of noise can confuse neural networks during the training phase. Therefore, it is important to properly adjust the noise power using in the training phase. Finally, to show the generalize ability of the Autoencoder to different block lengths, we use different block length in the testing phase, i.e. $L = 50$, show its performance. As can be seen, thanks to the weight-sharing property from the 1D-CNN model, identical SER performance is achieved with different block lengths. 
 
\vspace{-0.0cm}
Next, we evaluate the SER performance of the proposed approaches under different data modulation schemes. For the "Autoencoder", we use the noise power that provides the best SER performance for each modulation scheme for a fair comparison. In addition, we plot the SER performance with the proposed model-based approach as an benchmark\footnote{Although model-based approach does not directly optimize SER performance, it is known that the sum channel capacity maximization is equivalent to the minimum mean square error optimization which reduces SER at the receiver. Therefore, the model-based scheme is also expected to attain a low symbol error ratio as shown in Fig.~\ref{fig:SERCompare}.}. As can be seen in Fig.~\ref{fig:SERCompare}, both "Autoencoder" and "Model-based" approaches significantly improve the error rate of the detected symbol. Moreover, the "Autoencoder" outperforms the "Model-based" in both examined modulation schemes. The reason behind this is that the "Model-based" scheme optimizes the capacity of the system while the "Autoencoder" directly optimizes the error rate of the decoded symbols. Moreover, in the   "Model-based" approach, the fixed precoder and combiner are applied regardless of different transmitted symbols, while the "Autoencoder" approach could adaptively optimize the encoder, decoder, and RIS phase shift elements for different transmitting symbols. We also notice that the "Autoencoder" can encode the transmitting data and optimize the RIS phase shift elements without knowledge of CSI, which reduces the burden of feedback overhead in downlink channel estimation. 
\vspace{-0.2cm}
\section{Conclusion} \label{Sec:Conclusion}
In this paper, we investigated RIS-assisted MIMO communication systems over double-scattering channels with spatial correlation and finite scatterers under the presence of LoS components. First, we provided the upper bound for the channel power gain given the fixed RISs' reflecting coefficients. Then, we proposed an alternating optimization algorithm based on the ADMM method to optimize the RISs' phase shifts that improves system capacity. The optimal precoding and combining matrices were then obtained via the SVD method, given the optimal RISs' phase shifters. Moreover, we proposed an end-to-end framework to control the transceiver and RISs' reflecting elements to enhance communication reliability. In the proposed framework, we replaced each of the entities in the system with a one-dimension convolutional neural network. We jointly trained them to minimize the SER metric. Due to the cooperation between the transceiver and RISs'  reflecting elements, our framework showed superior enhancement in the data detection at the receiver.
\vspace{-0.2cm}
\appendix
\subsection{Proof of Lemma \ref{Lemma1}} \label{Appendix:Lemma1}
We start with providing the following lemma which plays a key role in this proof. 
\begin{lemma}[Lemma 8, \cite{van2020massive}] \label{UsefulLemma}
For a random matrix $\mathbf{G} \in \mathbb{C}^{M_1 \times M_2}$ with each element distributed as $\mathcal{CN}(0,1)$ and a positive-semidefinite matrix $\mathbf{U} \in \mathbb{C}^{N_2 \times N_2}$, it holds that
\begin{equation}
\mathbb{E} [ \mathbf{G} \mathbf{U} \mathbf{G}^H ] = \mathrm{tr}(\mathbf{U}) \mathbf{I}_{M_1}. 
\end{equation}
\end{lemma}

The second moment of the cascaded channel defined in \eqref{eq:H} is computed as
\begin{equation} \label{eq:EHH}
\begin{split}
    \mathbb{E}\left[\mathbf{O}\mathbf{O}^H \right] &= \mathbb{E}[(\mathbf{G}_2\mathbf{\Phi}_2\mathbf{D}\mathbf{\Phi}_1\mathbf{H}_1 + \mathbf{G}_1\mathbf{\Phi}_1\mathbf{H}_1 + \mathbf{G}_2\mathbf{\Phi}_2\mathbf{H}_2) \times \\
    &(\mathbf{G}_2\mathbf{\Phi}_2\mathbf{D}\mathbf{\Phi}_1\mathbf{H}_1 + \mathbf{G}_1\mathbf{\Phi}_1\mathbf{H}_1 + \mathbf{G}_2\mathbf{\Phi}_2\mathbf{H}_2)^H]\\
    &= \mathbb{E}[(\mathbf{A}_1+\mathbf{A}_2+\mathbf{A}_3)(\mathbf{A}_1+\mathbf{A}_2+\mathbf{A}_3)^H],
\end{split}
\end{equation}
where $\mathbf{A}_1 =\mathbf{G}_2\mathbf{\Phi}_2\mathbf{D}\mathbf{\Phi}_1\mathbf{H}_1$, $\mathbf{A}_2 = \mathbf{G}_1\mathbf{\Phi}_1\mathbf{H}_1$, and  $\mathbf{A}_3 = \mathbf{G}_2\mathbf{\Phi}_2\mathbf{H}_2)$. From its definition, we first compute the expectation $\mathbb{E}\{ \mathbf{A}_1 \mathbf{A}_1^H \}$ as follows
\begin{equation}
    \begin{split}
     &\mathbb{E}\big[\mathbf{A}_1\mathbf{A}_1^H\big] = \mathbb{E}\big[\mathbf{G}_2\mathbf{\Phi}_2\mathbf{D}\mathbf{\Phi}_1\mathbf{H}_1\mathbf{H}_1^H\mathbf{\Phi}_1^H\mathbf{D}^H\mathbf{\Phi}_2^H\mathbf{G}_2^H\big]\\
     &\stackrel{(a)}{=}\alpha_1\mathbb{E}\left[\mathbf{G}_2\mathbf{\Phi}_2\mathbf{D}\mathbf{\Phi}_1\mathbb{E}\left[\left(\sqrt{\frac{\epsilon_1}{\epsilon_1+1}}\mathbf{\bar{H}}_1+\sqrt{\frac{1}{\epsilon_1+1}}\mathbf{\tilde{H}}_1\right)\right.\right.\\
        &\left. \left. \left(\sqrt{\frac{\epsilon_1}{\epsilon_1+1}}\mathbf{\bar{H}}_1+\sqrt{\frac{1}{\epsilon_1+1}}\mathbf{\tilde{H}}_1\right)^H\right]\mathbf{\Phi}_1^H\mathbf{D}^H\mathbf{\Phi}_2^H\mathbf{G}_2^H\right ] \stackrel{(b)}{=} \alpha_1 \times 
\end{split}
\end{equation}
\begin{equation}
\begin{split}
     & \mathbb{E}\left[\mathbf{G}_2\mathbf{\Phi}_2\mathbf{D}\mathbf{\Phi}_1\left(\frac{\epsilon_1}{\epsilon_1+1}\mathbf{\bar{H}}_1\mathbf{\bar{H}}_1^H +\frac{1}{\epsilon_1+1}N_t\mathbf{R}_{t,\mathbf{H}_1}\right)\mathbf{\Phi}_1^H\mathbf{D}^H\mathbf{\Phi}_2^H\mathbf{G}_2^H\right], \notag
    \end{split}
\end{equation}
where $(a)$ is obtained by the definition of $\mathbf{H}_1$ in \eqref{eq:Hi} and the mutual independence between the propagation channels;   $(b)$ is obtained by the fact that $\mathbb{E}[\mathbf{\bar{H}}_1\mathbf{\tilde{H}}_1^H] = 0$ and $ \mathbb{E}\big[\mathbf{\tilde{H}}_1\mathbf{\tilde{H}}_1^H\big]$ is 
\begin{equation} \label{eq:EHH1tilde}
\begin{split}
     \mathbb{E}\big[\mathbf{\tilde{H}}_1\mathbf{\tilde{H}}_1^H\big] &\stackrel{(a)}{=} \frac{1}{\mathsf{SC}_{\mathbf{H}_1}}\mathbb{E}\left[\mathbf{R}_{t,\mathbf{H}_1}^{\frac{1}{2}}\mathbf{Q}_{\mathbf{H}_1}\mathbf{S}_{\mathbf{H}_1}^{\frac{1}{2}}\mathbf{P}_{\mathbf{H}_1}\mathbf{R}_{r,\mathbf{H}_1}\mathbf{P}_{\mathbf{H}_1}^H\mathbf{S}_{\mathbf{H}_1}^{\frac{1}{2}}\mathbf{Q}_{\mathbf{H}_1}^H\mathbf{R}_{t,\mathbf{H}_1}^{\frac{1}{2}}\right]\\
    &\stackrel{(b)}{=}\frac{\mathrm{tr}(\mathbf{R}_{r,\mathbf{H}_1})}{\mathsf{SC}_{\mathbf{H}_1}}\mathbb{E}\left[\mathbf{R}_{t,\mathbf{H}_1}^{\frac{1}{2}}\mathbf{Q}_{\mathbf{H}_1}\mathbf{S}_{\mathbf{H}_1}\mathbf{Q}_{\mathbf{H}_1}^H\mathbf{R}_{t,\mathbf{H}_1}^{\frac{1}{2}}\right]\\
    &\stackrel{(c)}{=} \frac{N_t\mathrm{tr}(\mathbf{S}_{\mathbf{H}_1})}{\mathsf{SC}_{\mathbf{H}_1}}\mathbf{R}_{t,\mathbf{H}_1} = \frac{N_t\mathrm{SC}_{\mathbf{H}_1}}{\mathrm{SC}_{\mathbf{H}_1}}\mathbf{R}_{t,\mathbf{H}_1} = N_t\mathbf{R}_{t,\mathbf{H}_1}.
\end{split}
\end{equation}
In \eqref{eq:EHH1tilde}, $(a)$ is obtained by utilizing the  double scattering model defined for $\tilde{\mathbf{H}}_1$ as in \eqref{eq:doubleScatter}; $(b)$ is obtained by exploiting Lemma~\ref{UsefulLemma} and the mutual independence of the small scale fading coefficients;  $(c)$ is obtained by noting that $\mathrm{tr}(\mathbf{R}_{r,\mathbf{H}_1}) = N_t$, $\mathrm{tr}(\mathbf{S}_{\mathbf{H}_1}) = \mathsf{SC}_{\mathbf{H}_1}$, and Lemma~\ref{UsefulLemma}. Following the same procedure, we can continue the computation as
\begin{align} \label{eq:EHHv1}
        &\mathbb{E}\big[\mathbf{A}_1\mathbf{A}_1^H\big] = \alpha_1\mathbb{E}\big[\mathbf{G}_2\mathbf{\Phi}_2\mathbf{D}\mathbf{X}\mathbf{D}^H\mathbf{\Phi}_2^H\mathbf{G}_2^H\big] \notag\\
        &= \alpha_1\gamma\mathbb{E}\left[\mathbf{G}_2\mathbf{\Phi}_2\left(\frac{\mu}{\mu+1}\mathbf{\bar{D}}\mathbf{X}\mathbf{\bar{D}}^H+\frac{1}{\mu+1}\mathbf{\tilde{D}}\mathbf{X}\mathbf{\tilde{D}}^H\right)\mathbf{\Phi}^H\mathbf{G}_2^H\right] \notag\\
        &= \alpha_1\gamma\mathbb{E}\left[\mathbf{G}_2\mathbf{\Phi}_2\left(\frac{\mu}{\mu+1}\mathbf{\bar{D}}\mathbf{X}\mathbf{\bar{D}}^H \right.\right. \notag\\
        &\left.\left. +\frac{1}{\mu+1}\mathrm{tr}(\mathbf{R}_{r,\mathbf{D}}^{\frac{1}{2}}\mathbf{X}\mathbf{R}_{r,\mathbf{D}}^{\frac{1}{2}})\mathbf{R}_{t,\mathbf{D}}\right)\mathbf{\Phi}_2^H\mathbf{G}_2^H\right]\notag\\
        &= \alpha_1\beta_2\gamma\frac{\mu \delta_2}{(\mu+1)(\delta_2+1)}\mathbf{\bar{G}}_2\mathbf{\Phi}_2\mathbf{\bar{D}}\mathbf{X}\mathbf{\bar{D}}^H\mathbf{\Phi}_2^H\mathbf{\bar{G}}_2^H \notag\\
        &+\alpha_1\beta_2\gamma\frac{\mu}{(\mu+1)(\delta_2+1)}\mathbb{E}[\mathbf{\tilde{G}}_2\mathbf{\Phi}_2\mathbf{\bar{D}}\mathbf{X}\mathbf{\bar{D}}^H\mathbf{\Phi}_2^H\mathbf{\tilde{G}}_2^H]\notag\\
        &+\alpha_1\beta_2\gamma\frac{\delta_2}{(\mu+1)(\delta_2+1)}\mathrm{tr}(\mathbf{R}_{r,\mathbf{D}}^{\frac{1}{2}}\mathbf{X}\mathbf{R}_{r,\mathbf{D}}^{\frac{1}{2}})\mathbf{\bar{G}}_2\mathbf{\Phi}_2\mathbf{R}_{t,\mathbf{D}}\mathbf{\Phi}_2^H\mathbf{\bar{G}}_2^H\notag\\
        &+ \alpha_1\beta_2\gamma\frac{1}{(\mu+1)(\delta_2+1)}\mathrm{tr}(\mathbf{R}_{r,\mathbf{D}}^{\frac{1}{2}}\mathbf{X}\mathbf{R}_{r,\mathbf{D}}^{\frac{1}{2}})\mathbb{E}[\mathbf{\tilde{G}}_2\mathbf{\Phi}_2\mathbf{R}_{t,\mathbf{D}}\mathbf{\Phi}_2^H\mathbf{\tilde{G}}_2^H]\notag\\
        &= \alpha_1\beta_2\gamma\frac{\mu \delta_2}{(\mu+1)(\delta_2+1)}\mathbf{\bar{G}}_2\mathbf{\Phi}_2\mathbf{\bar{D}}\mathbf{X}\mathbf{\bar{D}}^H\mathbf{\Phi}_2^H\mathbf{\bar{G}}_2^H \notag\\
        &+\alpha_1\beta_2\gamma\frac{\mu}{(\mu+1)(\delta_2+1)}\mathrm{tr}(\mathbf{R}_{r,\mathbf{G}_2}^{\frac{1}{2}}\mathbf{\Phi}_2\mathbf{\bar{D}}\mathbf{X}\mathbf{\bar{D}}^H\mathbf{\Phi}_2^H\mathbf{R}_{r,\mathbf{G}_2}^{\frac{1}{2}})\mathbf{R}_{t,\mathbf{G}_2}\notag\\
        &+ \alpha_1\beta_2\gamma\frac{\delta_2}{(\mu+1)(\delta_2+1)}\mathrm{tr}(\mathbf{R}_{r,\mathbf{D}}^{\frac{1}{2}}\mathbf{X}\mathbf{R}_r,{\mathbf{D}}^{\frac{1}{2}})\mathbf{\bar{G}}_2\mathbf{\Phi}_2\mathbf{R}_{t,\mathbf{D}}\mathbf{\Phi}_2^H\mathbf{\bar{G}}_2^H \notag\\
        &+ \alpha_1\beta_2\gamma\frac{1}{(\mu+1)(\delta_2+1)}\mathrm{tr}(\mathbf{R}_{r,\mathbf{D}}^{\frac{1}{2}}\mathbf{X}\mathbf{R}_{r,\mathbf{D}}^{\frac{1}{2}})\notag\\
        &\times \mathrm{tr}(\mathbf{R}_{r,\mathbf{G}_2}^{\frac{1}{2}}\mathbf{\Phi}_2\mathbf{R}_{t,\mathbf{D}}\mathbf{\Phi}_2^H\mathbf{R}_{r,\mathbf{G}_2}^{\frac{1}{2}})\mathbf{R}_{t,\mathbf{G}_2},
\end{align}
where $\mathbf{X} = \mathbf{\Phi}_1(\frac{\epsilon_1}{\epsilon_1+1}\mathbf{\bar{H}}_1\mathbf{\bar{H}}_1^H+\frac{1}{\epsilon_1+1}N_t\mathbf{R}_{\mathbf{H}_1})\mathbf{\Phi}_1^H$. 
By applying the methodology to the other expectations in \eqref{eq:EHH}, we can  obtain the result in the lemma.
\vspace{-0.3cm}

\bibliographystyle{IEEEtran}
\bibliography{refs}
%\bibliographystyle{IEEEtran}
%\bibliography{IEEEabrv,refs}
\begin{IEEEbiography}     [{\includegraphics[width=1in,height=1.2in,clip,keepaspectratio]{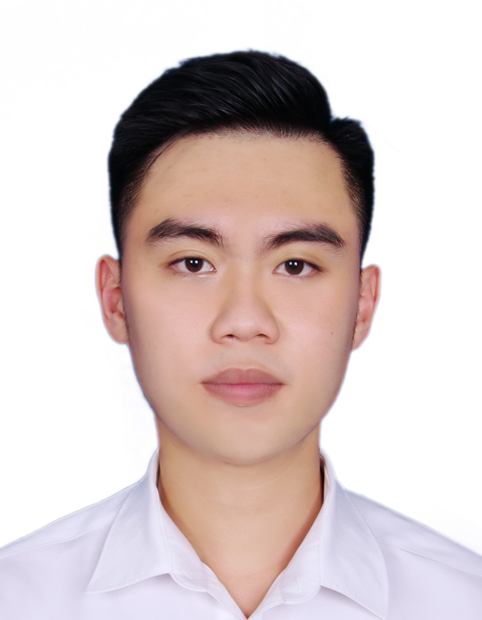}}]
	{Ha An Le} received the B.S. degree in electronic and telecommunication from Hanoi University of Science and Technology, Hanoi, Vietnam in 2020. He is currently pursuing the combined Masters/Doctorate degree in the Department of Electrical and Computer Engineering, Seoul National University, working in the Wireless Communication and Information Systems Laboratory directed by Prof. Wan Choi. His current research interests include wireless communication and machine learning.

\end{IEEEbiography}
	\begin{IEEEbiography} 
    [{\includegraphics[width=1.0in,height=1.25in,clip,keepaspectratio]{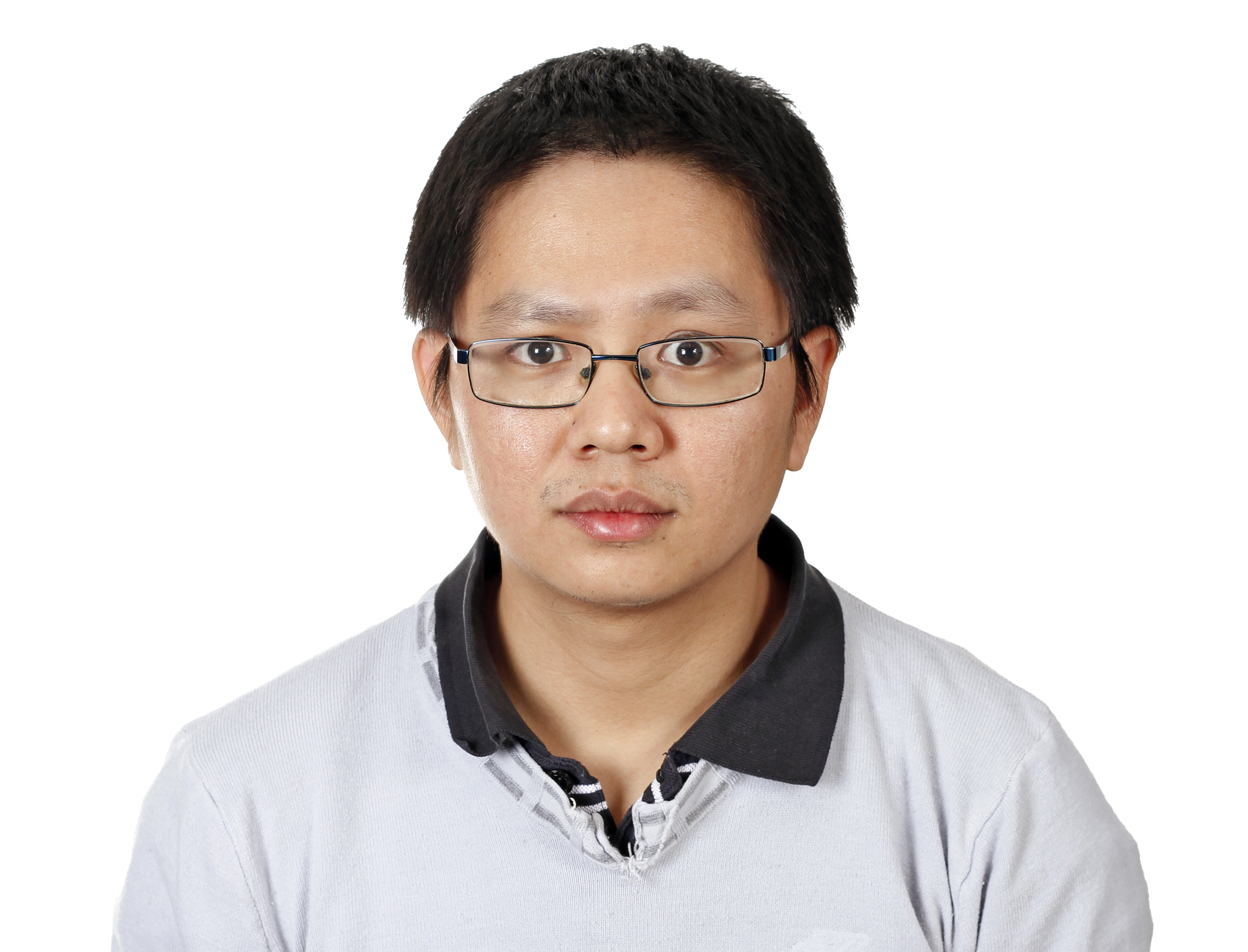}}]
	{Trinh Van Chien} (S'16-M'20) received the B.S. degree in Electronics and Telecommunications from Hanoi University of Science and Technology (HUST), Vietnam, in 2012. He then received the M.S. degree in Electrical and Computer Engineering from Sungkyunkwan University (SKKU), Korea, in 2014 and the Ph.D. degree in Communication Systems from Link\"oping University (LiU), Sweden, in 2020. He was  a research associate at the University of Luxembourg. He is now with the School of Information and Communication Technology (SoICT), Hanoi University of Science and Technology (HUST), Vietnam. His interest lies in convex optimization problems and machine learning applications for wireless communications and image \& video processing. He was an IEEE Wireless Communications Letters exemplary reviewer for 2016, 2017, and 2021. He further received an IEEE Transactions on Communications exemplary reviewer for 2022. He also received the award of scientific excellence in the first year of the 5Gwireless project funded by European Union Horizon's 2020. He  co-received the best student paper award in IEEE ATC 2022. 
\vspace{-1cm}
\end{IEEEbiography}
\begin{IEEEbiography}     [{\includegraphics[width=1.0in,height=1.2in,clip,keepaspectratio]{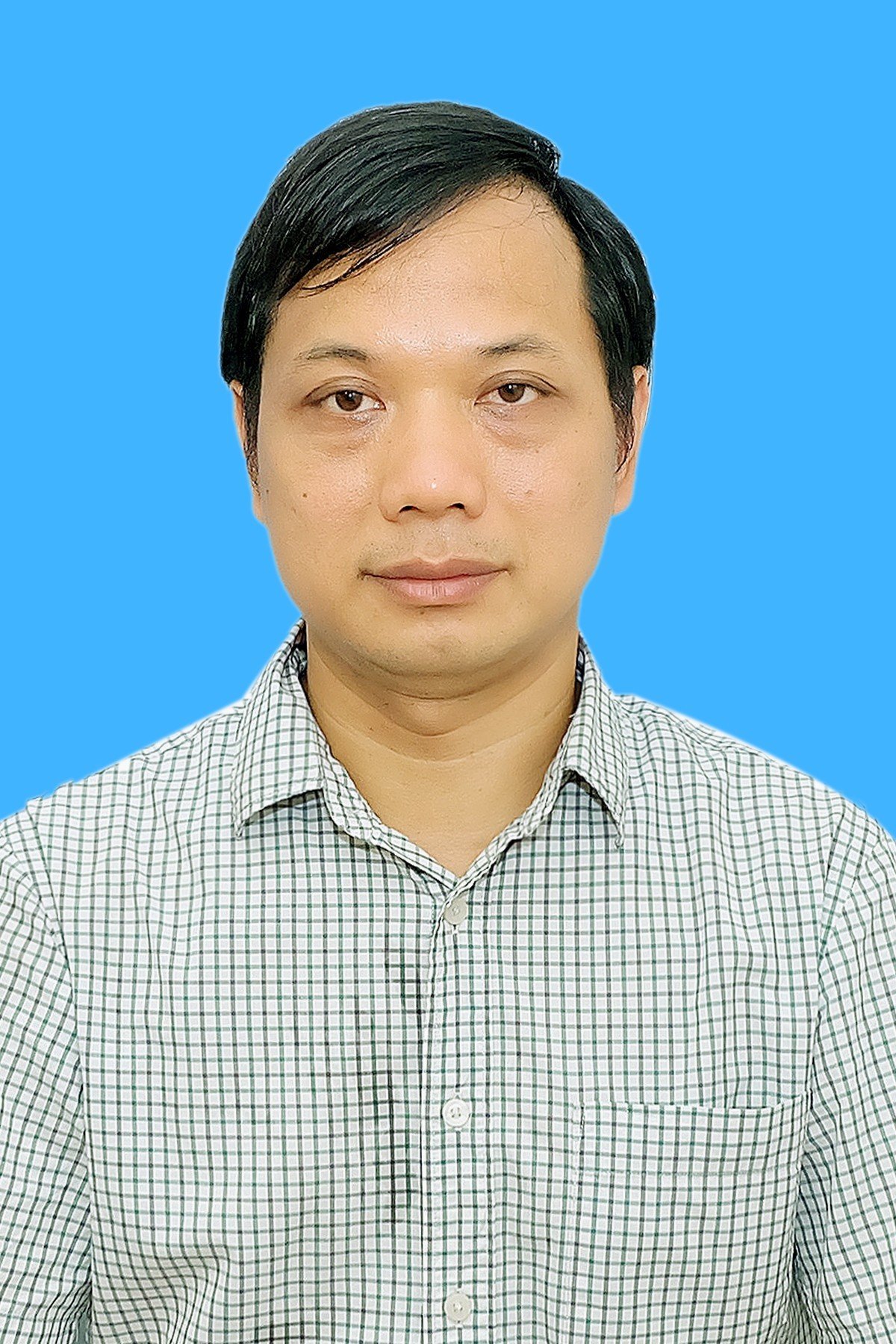}}]
	{Van Duc Nguyen} received the Bachelor and  Master of Engineering degrees in electronics and communications from Hanoi University of  Science and Technology, Vietnam, in 1995 and 1997, respectively, and the Dr. Eng. degree in communications engineering from University of Hannover, Germany, in 2003.  From 1998 to 2003, he was with the Institute of Communications Engineering, University of Hannover, first as a DAAD Scholarship Holder and then as a member of the scientific staff. From 2003 to 2004, he was a Post-Doctoral Researcher at Agder University College, Grimstad, Norway. From 2005 to 2006, he worked at the International University of Bremen as a Post-Doctoral Fellow. In 2007, he spent two months at  Sungkyungkwan University, South Korea, as a Research Professor. In 2008, he was a Visiting Researcher at Klagenfurt University, Austria. In 2009, he spent one month at Agder University, Norway, as a Visiting Researcher. His current research interests include mobile radio communications, especially MIMO-OFDM systems, and radio resource management, channel coding for cellular and ad-hoc networks.

\end{IEEEbiography}
\begin{IEEEbiography} 
[{\includegraphics[width=1.0in,height=1.25in,clip,keepaspectratio]{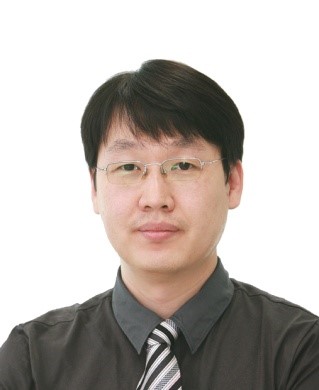}}]
    {Wan Choi} (S’03–M’06–SM’12-F’20) is Professor of Department of Electrical and Computer Engineering, Seoul National University, Seoul, Korea. From Feb. 2007 to Feb. 2020, he was Professor of School of Electrical Engineering, Korea Advanced Institute of Science and Technology (KAIST), Daejeon, Korea. He received the B.Sc. and M.Sc. degrees from the School of Electrical Engineering and Computer Science (EECS), Seoul National University (SNU), Seoul, Korea, in 1996 and 1998, respectively, and the Ph.D. degree in the Department of Electrical and Computer Engineering at the University of Texas at Austin in 2006. From 1998 to 2003, he was a Senior Member of the Technical Staff of the R\&D Division of KT, Korea, where he researched 3G CDMA systems. 
    He is the recipient of IEEE Vehicular Technology Society Jack Neubauer Memorial Award (Best System Paper Award) in 2002. He also received the IEEE Vehicular Technology Society Dan Noble Fellowship Award in 2006, the IEEE Communication Society Asia Pacific Young Researcher Award in 2007, Haedong Young Scholar Award from KICS in 2012, and the Irwin-Jacobs Award from Qualcomm and KICS in 2015. While at the University of Texas at Austin, he was the recipient of William S. Livingston Graduate Fellowship and Information and Telecommunication Fellowship from Ministry of Information and Communication (MIC), Korea. He is an Area Editor for the IEEE Transactions on Wireless Communications from Aug. 2022 and an Editor for the IEEE Transactions on Vehicular Technology from Apr. 2011. He served as the Executive Editor Chair for the IEEE Transactions on Wireless Communications (2019- 2021) and Executive Editor (2014-2019). He was also an Editor for the IEEE Transactions on Wireless Communications (2009-2014), for the IEEE Wireless Communications Letter (2012-2017), and as Guest Editor for the IEEE Journal on Selected Areas in Communications.
\end{IEEEbiography}
\end{document}